\documentclass[12pt]{article}
\usepackage{latexsym,amsmath,amsfonts,amssymb}
\usepackage[latin1]{inputenc}
\usepackage[american]{babel}
\usepackage[dvips]{graphicx}
\usepackage{bbm}
\newcommand{\dvol}{d\mbox{vol}}

\newcommand{\Nugual}[1]{$\mathcal{N}= #1 $}

\newcommand{\parfrac}[2]{\frac{\partial #1}{\partial #2}}
\newcommand{\delfrac}[2]{\frac{\delta #1}{\delta #2}}
\newcommand{\orders}[1]{\calO \Bigl( #1 \Bigr)}
\newcommand{\be}{\begin{equation}} \newcommand{\ee}{\end{equation}}
\newcommand{\bea}{\begin{equation} \begin{aligned}} \newcommand{\eea}{\end{aligned} \end{equation}}
\newcommand{\tabs}{\rule[-1ex]{0pt}{3.5ex}}

\numberwithin{equation}{section}

\newcommand{\Tr}{\mbox{Tr}}    % trace over gauge indices

\newcommand{\calF}{\mathcal{F}}
\newcommand{\calO}{\mathcal{O}}
\newcommand{\calM}{\mathcal{M}}

\newcommand{\calC}{\mathcal{C}}
\newcommand{\calR}{\mathcal{R}}
\newcommand{\bbZ}{\mathbb{Z}}
\newcommand{\bbC}{\mathbb{C}}
\newcommand{\bbP}{\mathbb{P}}

\newcommand{\bbR}{\mathbb{R}}

\begin{document}

\begin{titlepage}
\rightline{SISSA 67/2007/EP}
\vspace{0.3in}

\begin{center}
{\LARGE \bf A Chiral Cascade via Backreacting \\[.3cm] D7-branes with Flux}
\end{center}
\vskip 0.6truein
\begin{center}
{\Large Francesco Benini}      %${}^{*}$\footnote{benini@sissa.it}
\vspace{0.4in}\\
%${}^{*}$
\it{ SISSA/ISAS and INFN-Sezione di Trieste\\ Via Beirut 2;
I-34014 Trieste, Italy \\
email address: benini@sissa.it}
\end{center}
\vspace{2.5cm}
%\begin{center}
\centerline{{\bf Abstract}}
\vspace{.5cm}

%{\setlength{\baselineskip}{1.2\baselineskip}
{\setlength{\baselineskip}{\baselineskip}
% In the context of AdS/CFT, a cascading theory with an arbitrary number of chiral flavors is considered. On the gravity side, the theory is engineered with intersecting D7-branes with world-volume flux put on a conifold with 3-form fluxes. New fully backreacted solutions of Type IIB Supergravity plus branes are found.

In the context of AdS/CFT, we consider a cascading theory with an arbitrarily large number of chiral flavors. In the UV the theory can be considered as a chiral flavoring of the Klebanov-Tseytlin solution, and exhibits a duality wall. Instead in the IR, due to the rich dynamics, it safely flows to a non-cascading theory. We engineer the field theory through intersecting D7-branes with world-volume gauge flux on a conifold with 3-form fluxes, and we find new fully backreacted solutions of Type IIB Supergravity plus branes. We match the field theory cascade with supergravity by computing Page charges and interpreting Seiberg dualities as large gauge transformations of the background. Eventually we give an interpretation of the chiral zero modes arising at the intersection of the D7-branes with flux.
\par}

% New analytic solutions of Type IIB Supergravity with backreacting intersecting D7-branes
%
%
%
% I compute the backreaction of D7-branes with flux ... chiral cascade ... mathcing the cascade with supergravity ... Page charges and Seiberg duality as a large gauge transformation ... chiral zero modes originating from intersecting D7-branes with flux.\par}
\end{titlepage}
\setcounter{footnote}{0}
%\tableofcontents
%--------+---------+---------+---------+---------+---------+---------+
%Body

\newpage

%\tableofcontents

\section{Introduction}

Early ideas of t'Hooft \cite{'t Hooft:1973jz} suggested that the physics and dynamics of strong interactions could be understood and described through a theory of strings. Such an idea started materializing with the advent of Maldacena's conjecture \cite{Maldacena:1997re}, also known as AdS/CFT. The authors showed that a theory of strings can capture both perturbative and non-perturbative aspects of a (3+1) dimensional field theory. Unfortunately the field theory in question, namely \Nugual{4} super Yang-Mills (SYM), does not have great phenomenological interest. A first step towards phenomenologically more relevant generalizations was made by extending the AdS/CFT duality to branes at conical singularities \cite{Acharya:1998db,Klebanov:1998hh}. The prototype example is that of D3-branes at a conifold singularity (see, among the more relevant papers, \cite{Klebanov:1998hh, Gubser:1998fp, Klebanov:1999rd, Klebanov:2000nc, Klebanov:2000hb, Gubser:2004qj, Dymarsky:2005xt, Butti:2004pk, Strassler:2005qs}).
This improvement provided a way of breaking the large amount of supersymmetry to the minimal one, and breaking the conformal symmetry as well.

A characteristic of all the models realized with branes at singularities is that the dual field theories only contain fields in the adjoint or bifundamental representation of the gauge factors. Obviously for phenomenological reasons the next step is the inclusion of matter in the fundamental representation. A beautiful example appeared in \cite{Karch:2002sh}, where the new degrees of freedom where introduced in the brane picture through extra non-compact flavor branes.

The difference between color and flavor branes is substantial. Color branes undergo a geometric transition and `disappear': the open string dynamics on them is equally described by a dual closed string background with fluxes. Flavor branes instead are still present into the dual background after the geometric transition. Being non-compact, they do not have a 4d gauge dynamics. On the other hand they do have an higher dimensional gauge theory living on them which, according to the usual AdS/CFT dictionary, is dual to a global symmetry in field theory. In an appropriate large $N_c$ and small $g_s$ regime, the system can be described in supergravity; the action must however be enriched with a Dirac-Born-Infeld and Wess-Zumino piece to describe the flavor branes:
\begin{equation}
S = S_{IIB} + S_{DBI} + S_{WZ} \;.
\end{equation}

Many ideas and examples originated from the previous setup \cite{Kruczenski:2003be,Ouyang:2003df,Kuperstein:2004hy}. All those frameworks are good for a regime where the number of flavors $N_f$ is much smaller than the number of colors $N_c$, because they restrict to the so called \emph{quenched approximation}. They deal with probe flavor branes that do not backreact on the closed string sector. From a field theoretical diagrammatic point of view, they give the correct physics in the t'Hooft limit with $N_f = fixed$ \cite{'t Hooft:1973jz}, where quarks are only external legs not participating in loops.

On the other hand some papers appeared where the backreaction of the flavor branes is taken into account, realizing in this way the Veneziano expansion with $N_f/N_c = fixed$ \cite{Veneziano:1976wm}. Early works are \cite{Grana:2001xn,Bertolini:2001qa}. More recently a series of papers appeared where the backreaction of the flavor branes is handled through a powerful smearing technique.
In \cite{Casero:2006pt} it was proposed a gravity dual to \Nugual{1} SQCD (see also \cite{Casero:2007jj}), in \cite{Paredes:2006wb} to \Nugual{2} SQCD and in \cite{Benini:2006hh} to the Klebanov-Witten (KW) theory \cite{Klebanov:1998hh} with flavors.
In \cite{Benini:2007gx} it was considered the addition of new degrees of freedom to the Klebanov-Tseytlin (KT) \cite{Klebanov:2000nc} and Klebanov-Strassler (KS) \cite{Klebanov:2000hb} setups. The new degrees of freedom are non-chiral fundamental matter, which enriches a lot the duality cascade of the renormalization group (RG) flow.

In this paper I extend the smearing technique to a case of chiral fundamental matter. The new ingredient is that the flavor branes needed to realize such a field theory have a non-trivial gauge bundle on them. First of all taking into account the flux raises new issues about supersymmetry. Then the gauge flux induces new charges which have to be taken into account, and it could give rise to new modes at the intersection of flavor branes. The paper can thus be thought as a generalization of the smearing technique to the case of non-trivial gauge bundles. The interest resides in the fact that the chiral case is much more generic than the non-chiral one, when one tries to extend the flavoring of cascading theories done in \cite{Benini:2007gx} to fractional branes at more generic conical singularities. The non-chiral case (flavor branes with trivial gauge bundle) seems to be quite special.

The KT and KS supergravity solutions have a field theory dual whose RG flow can be understood as a cascade of Seiberg dualities. When the ranks of the gauge factors are different, they reduce along the flow while their difference remains constant. In presence of flavors also the difference reduces along the cascade \cite{Benini:2007gx}, possibly reaching an IR theory with equal ranks that does not cascade any more. This is the flow considered in this paper. Moreover in the chiral case there is a further issue: along the cascade new gauge singlet fields appear/disappear, in order to match a global anomaly. They have a beautiful interpretation as chiral zero modes living at the intersection of flavor branes with flux. The existence of these modes was already noticed in \cite{Franco:2006es}, where the authors used a similar chiral cascade to realize ISS vacua \cite{Intriligator:2006dd} at the bottom.

Following the ideas of \cite{Benini:2007gx}, Seiberg dualities are interpreted in supergravity as large gauge transformations. This gets nicely married with the fact that gauge ranks are measured by Page charges, rather than Maxwell charges. In this paper for the first time there is a disagreement between the two, and only Page charges give an exact matching with field theory.

The paper is organized as follows. In Section 2 I analyze the field theory with chiral matter and the cascade describing its RG flow. In Section 3 I construct a supergravity dual in the $N_f \ll N_c$ limit with probe D7-branes and I show that a non-trivial gauge bundle is needed. In Section 4 I present the supergravity equations with sources, and I use them in Section 5 to construct a backreacted solution for the case $N_f \sim N_c$. In Section 7 I compute Maxwell and Page charges; I also give an holographic interpretation of the gauge singlet fields. In Section 8 I construct a dictionary between supergravity and field theory, and a perfect matching is verified. Conventions and computations are in various Appendices.

\section{A Field Theory Cascade}

\begin{figure}[t]
\begin{minipage}{0.45\textwidth}
\begin{center}
\includegraphics[height=0.63\textwidth]{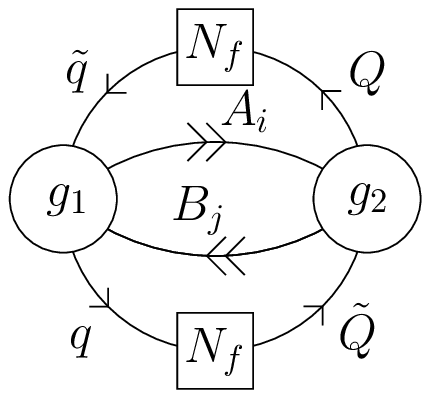}
\caption{Quiver diagram of the electric theory. \label{fig:electric theory1}}
\end{center}
\end{minipage}
\hspace{\stretch{1}}
\begin{minipage}{0.45\textwidth}
\begin{center}
\includegraphics[height=0.63\textwidth]{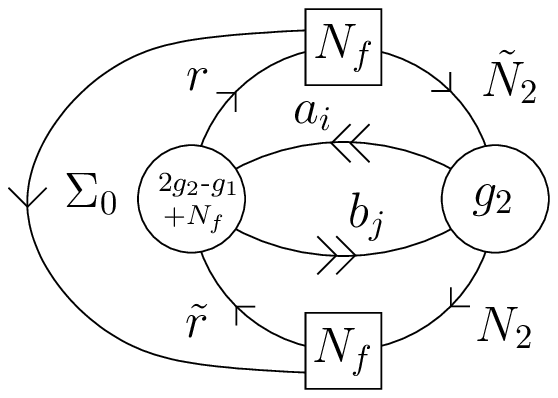}
\caption{Quiver diagram of the magnetic theory after Seiberg duality. \label{fig:magnetic theory1}}
\end{center}
\end{minipage}
\end{figure}

Consider a field theory whose quiver diagram is depicted in Figure \ref{fig:electric theory1}. It consists of two gauge groups $SU(g_1) \times SU(g_2)$ (where for definiteness we take $g_1 > g_2$) and two flavor groups $U(N_f) \times U(N_f)$. Part of the flavor group is generically anomalous: the axial $U(1)_{fA}$ always has a flavor-gauge-gauge triangle anomaly, while for $g_1 \neq g_2$ both $U(N_f)$ factors have a flavor-flavor-flavor anomaly with only the diagonal $U(N_f)_{fV}$ anomaly-free. There are four bifundamental fields $A_i$ and $B_i$ with $i=1,2$ and four (anti)fundamental fields $q$, $\tilde q$, $Q$, $\tilde Q$.
The superpotential I consider is
\begin{equation} \label{superpotential Ouyang}
W = h\, (A_1B_1A_2B_2 - A_1B_2A_2B_1) + \lambda\, (\tilde qA_1Q + \tilde QB_1q)
\end{equation}
where traces on color and flavor indices are meant. The $SU(2)\times SU(2)$ flavor symmetry of the theory without fundamental fields (acting on $A_i$ and $B_j$) is broken to its toric subgroup by the superpotential. Moreover there are two baryonic symmetries $U(1)_B$ (which actually is the diagonal $U(1)$ subgroup of the flavor group) and $U(1)_{B'}$ and an anomalous R-symmetry $U(1)_R$. The theory is chiral, in the sense that we cannot construct mass terms without breaking the flavor symmetry. All the relevant charges are summarized in Table \ref{tab:charges}.

\begin{table}[t]
\begin{center}
\begin{tabular}{|r|c|c|c|c|c|c|}
\hline \tabs
& $SU(r_1)\times SU(r_2)$&$U(N_f)\times U(N_f)$&$SU(2)^2$&$U(1)_R$& $U(1)_B$ &$U(1)_{B'}$ \\
\hline\hline \tabs
$A_i$ & $(r_1,\overline{r_2})$ & $(1,1)$              & $(2,1)$   & $1/2$  & $1$ & $0$ \\
\tabs
$B_i$ & $(\overline{r_1},r_2)$ & $(1,1)$              & $(1,2)$   & $1/2$    & $-1$ & $0$ \\
\tabs
$q$   & $(r_1,1)$                & $(1,\overline{N_f})$   & $(1,1)$   & $3/4$ & $1$ & $1$ \\
\tabs
$\tilde q$ & $(\overline{r_1},1)$ & $(N_f,1)$           & $(1,1)$ & $3/4$ & $-1$ & $-1$ \\
\tabs
$Q$        & $(1,r_2)$            & $(\overline{N_f},1)$   & $(1,1)$  & $3/4$ & $0$ & $1$ \\
\tabs
$\tilde Q$ & $(1,\overline{r_2})$     & $(1,N_f)$          & $(1,1)$ & $3/4$ & $0$ & $-1$ \\
\tabs
$\tilde\Phi_k$ & $(1,1)$ & $(N_f,\overline{N_f})$ & $(1,1)$ & $\frac{1}{2}-k$ & $0$ & $0$ \\
\tabs
$\Phi_k$   & $(1,1)$     & $(\overline{N_f},N_f)$ & $(1,1)$ & $\frac{1}{2}-k$ & $0$ & $0$ \\
\hline
\end{tabular}
\caption{Field content and symmetries of the chirally flavored KT theory. \label{tab:charges}}
\end{center}
\end{table}

% \begin{table}[ht]
% \begin{center}
% \begin{tabular}{r|cccccc}
% %\hline
%            & $SU(g_1) \times SU(g_2)$ & $U(N_f) \times U(N_f)$ & $SU(2)^2$ & $U(1)_R$ & $U(1)_B$ & $U(1)_{B'}$ \\
% \hline
% $A_i$      & $(g_1,\overline{g_2})$   & $(1,1)$                & $(2,1)$   & $1/2$    & $0$      & $1$         \\
% $B_i$      & $(\overline{g_1},g_2)$   & $(1,1)$                & $(1,2)$   & $1/2$    & $0$      & $-1$        \\
% $q$        & $(g_1,1)$                & $(1,\overline{N_f})$   & $(1,1)$   & $3/4$    & $1$      & $1$         \\
% $\tilde q$ & $(\overline{g_1},1)$     & $(N_f,1)$              & $(1,1)$   & $3/4$    & $-1$     & $-1$        \\
% $Q$        & $(1,g_2)$                & $(\overline{N_f},1)$   & $(1,1)$   & $3/4$    & $1$      & $0$         \\
% $\tilde Q$ & $(1,\overline{g_2})$     & $(1,N_f)$              & $(1,1)$   & $3/4$    & $-1$     & $0$         \\
% $\tilde\Phi_k$ & $(1,1)$              & $(N_f,\overline{N_f})$ & $(1,1)$   & $\frac{1}{2}-k$ & $0$ & $0$       \\
% $\Phi_k$   & $(1,1)$                  & $(\overline{N_f},N_f)$ & $(1,1)$   & $\frac{1}{2}-k$ & $0$ & $0$
% %\hline
% \end{tabular}
% \caption{Field content and symmetries of the chirally flavored KT theory. \label{tab:charges}}
% \end{center}
% \end{table}

The theory without flavors and with $g_1=g_2$ has a complex line of conformal points \cite{Klebanov:1998hh,Benvenuti:2005wi}, where the anomalous dimensions can be derived from the non-anomalous R-charges.
If we take the number of colors $g_1$ and $g_2$ much larger than the number of flavors $N_f$, and we suppose that the anomalous dimensions of the bifundamentals only take corrections at second order in $N_f/N_c$ (this hypothesis was supported by a dual gravity analysis in \cite{Ouyang:2003df,Benini:2006hh}), we can compute the NSVZ gauge $\beta$-functions \cite{Novikov:1983uc}:
\begin{equation} \label{beta functions}
\beta_{G_1} = - \frac{3G_1^3}{16\pi^2} \Bigl[ g_1-g_2 - \frac{N_f}{4} \Bigr] \qquad\qquad
\beta_{G_2} = \frac{3G_2^3}{16\pi^2} \Bigl[ g_1 - g_2 + \frac{N_f}{4} \Bigr] \;.
\end{equation}
We find that, if the difference $(g_1-g_2)$ is larger than $N_f/4$, in the IR $SU(g_1)$ flows to strong coupling while $SU(g_2)$ flows to weak coupling. We can then perform a Seiberg duality \cite{Seiberg:1994pq} on node $SU(g_1)$. The mesons are: $B_i A_j \equiv M_{ij}$, $\tilde q A_i \equiv \tilde N_i$, $B_i q \equiv N_i$, $\tilde q q \equiv \Sigma_0$. The superpotential in the magnetic theory is
\begin{equation} \begin{aligned}
W' &= h\, (M_{12}M_{21} - M_{11}M_{22}) + \lambda\, (\tilde N_1Q + \tilde QN_1) + \frac{1}{\hat\Lambda}\, [a_j b_i M_{ij} + a_i r \tilde N_i + \tilde r b_i N_i + \tilde r r \Sigma_0 ] \;,
\end{aligned} \end{equation}
where we sum over $i,j=1,2$. $\hat\Lambda$ is the dynamically generated scale involved in Seiberg duality \cite{Seiberg:1994pq}, and represents the energy scale where we transit from a good electric description to a good magnetic description. Then we integrate out $M_{ij}$, $N_1$, $\tilde N_1$, $Q$, $\tilde Q$. The relevant F-term equations are:
\begin{equation} \begin{aligned} \label{F-term eqs}
-h\, M_{22} + \frac{1}{\hat\Lambda}\, a_1b_1 &= 0 \\
h\, M_{21} + \frac{1}{\hat\Lambda}\, a_2b_1 &= 0
\end{aligned} \qquad\qquad \begin{aligned}
\lambda\, Q + \frac{1}{\hat\Lambda}\, a_1 r &= 0 \\
\lambda\, \tilde Q + \frac{1}{\hat\Lambda}\, \tilde r b_1 &= 0 \;,
\end{aligned} \end{equation}
so that we obtain
\begin{equation}
W' = \frac{1}{h\hat\Lambda^2}\, (a_1b_1a_2b_2 - a_1b_2a_2b_1) + \frac{1}{\hat\Lambda}\, (\tilde N_2 a_2 r + \tilde r b_2 N_2 + \Sigma_0 \tilde r r ) \;.
\end{equation}
Notice that the mesonic fields have non-canonical mass dimension 2, and after canonical normalization of all fields some order one coupling constants could arise from the K\"ahler potential.

The magnetic quiver is depicted in Figure \ref{fig:magnetic theory1}. To compare it with the original electric quiver, we relabel the fields: $a_i,b_j \to A_i,B_j$ exchanging $1 \leftrightarrow 2$; $r,\tilde r \to Q,\tilde Q$ and $N_2, \tilde N_2 \to q,\tilde q$; recall that the biggest rank is now $g_2$. We see that the theory has reproduced itself, apart from the new gauge singlet field $\Sigma_0$ in the $(N_f, \overline{N_f})$ representation of the flavor group and a shift in gauge ranks: $(g_1,g_2) \to (g_2,2g_2 - g_1 + N_f)$. Even the superpotential has reproduced itself, with the quarks coupling with $A_1$ and $B_1$, apart from the new superpotential term $\frac{1}{\hat \Lambda} \Sigma_0 \tilde Q Q$.
Notice that the gauge singlet $\Sigma_0$ is there because of ``conservation'' of the global flavor-flavor-flavor anomaly of the axial $U(N_f)_{fA}$. In particular, the axial $U(1)_{fA}$ is broken by the anomaly to $\bbZ_{g1-g2}$, and this is true both in the electric and magnetic quiver. The dual gravity interpretation of this will be discussed in Section \ref{sec:brane engineering}.

Now let me ask what is the fate of the gauge singlet field. The theory continues flowing in the IR until another Seiberg duality is required. So we can generically consider a theory as in Figure \ref{fig:electric theory1} but with an extra gauge singlet $\tilde \Phi_k$ in the $(N_f, \overline{N_f})$ flavor representation from the beginning, and superpotential
\begin{equation}
W = h\, (A_1B_1A_2B_2 - A_1B_2A_2B_1) + \lambda\, (\tilde qA_1Q + \tilde QB_1q) + \lambda_k \, \tilde\Phi_k \, \tilde Q (B_2A_2)^k Q \;,
\end{equation}
not summed over $k$. As will be clear momentarily, it is better to consider a general superpotential depending on $k$, even if here we are interested in $k=0$. We perform a Seiberg duality on node $SU(g_1)$ as before, and integrate out $M_{ij}$, $N_1$, $\tilde N_1$, $Q$, $\tilde Q$ (the F-term equations are still \eqref{F-term eqs}). We obtain
\begin{equation}
W' = \frac{1}{h\hat\Lambda^2}\, (a_1b_1a_2b_2 - a_1b_2a_2b_1) + \frac{1}{\hat\Lambda}\, (\tilde N_2 a_2 r + \tilde r b_2 N_2 + \Sigma_0 \tilde r r) + \frac{\lambda_k}{h^k \lambda^2 \hat\Lambda^{k+2}}\, \tilde\Phi_k \, \tilde r (b_1a_1)^{k+1}r \;.
\end{equation}

We learn that at each Seiberg duality a new gauge singlet field in the $(N_f, \overline{N_f})$ representation is generated, while the existing ones develop longer and longer superpotential terms. We can try to estimate the behavior of the superpotential terms $\tilde \calO_k = \tilde \Phi_k \, \tilde Q (B_2A_2)^k Q$ under the RG flow. We consider again a regime of parameters where $g_1$ and $g_2$ are much larger than $(g_1-g_2)$ and $N_f$, so that the theory is close to its conformal points. Then the quantum dimensions of the fields $A$, $B$, $q$, $\tilde q$, $Q$, $\tilde Q$ can be derived from the R-charges (see Table \ref{tab:charges}) through the relation $D[\calO] = \frac{3}{2} R_\calO$, strictly valid at a conformal point. From the supergravity computation of the gauge coupling $\beta$-functions and their matching with field theory, one deduces that the quantum dimensions of $A$ and $B$ take corrections of order $(N_f/N_c)^2$, whilst the quark field ones of order $N_f/N_c$ \cite{Ouyang:2003df,Benini:2006hh}. The gravity computation does not tell us nothing about the quantum dimension of $\tilde\Phi_k$ since it does not enter in the $\beta$-functions, and in fact the dimension must take corrections of order one. Recall that gauge singlet scalars must have quantum dimension bigger than or equal to $1$, around a conformal point. We conclude that the superpotential terms $\tilde \calO_k = \tilde \Phi_k \, \tilde Q (B_2A_2)^k Q$ not only are irrelevant (their quantum dimension is bigger than $3$), but become more and more irrelevant going towards the IR (their quantum dimension runs). Notice that the fields $\tilde \Phi_k$ always couple with the quarks of the smallest gauge group.

Apart from this, the theory reproduces itself and cascades down, with both the ranks and the difference of the gauge group ranks reducing. From this point of view this chiral theory is similar to the one studied in \cite{Benini:2007gx}, but with the important difference that in this one $(g_1 - g_2)$ scales by $N_f$, while in the latter it scales by $N_f/2$. We will match this behavior with the dual gravity description in Section \ref{sec:brane engineering}.

\

\begin{figure}[t]
\begin{center}
\includegraphics[height=0.28\textwidth]{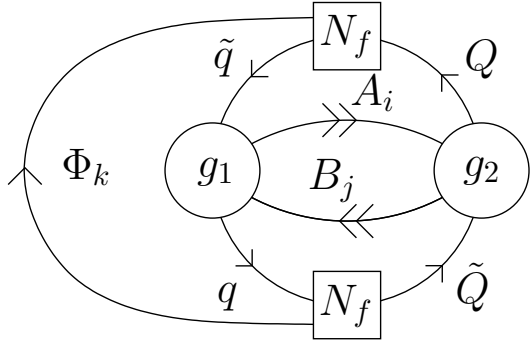}
\caption{Quiver diagram of an electric theory with a gauge singlet field $\Phi_k$ in the $(\overline{N_f},N_f)$ flavor representation. \label{fig:electric theory2}}
\end{center}
\end{figure}

Last but not least, I want to understand what happens if we start with a gauge singlet field $\Phi_k$ in the opposite flavor representation: $(\overline{N_f},N_f)$ (this implies that it couples to the quarks of the largest gauge group). The quiver is in Figure \ref{fig:electric theory2}. I will consider two cases at the same time: with minimal superpotential and with a larger one.
\begin{equation}
W = h\, (A_1B_1A_2B_2 - A_1B_2A_2B_1) + \lambda\, (\tilde qA_1Q + \tilde QB_1q) + \alpha_0\, \Phi_0 \tilde q q + \alpha_k\, \Phi_k \tilde q(A_2B_2)^k q \;.
\end{equation}
We perform a Seiberg duality going to the magnetic description as before:
\begin{equation} \begin{aligned}
W' &= h\, (M_{12}M_{21} - M_{11}M_{22}) + \lambda (\tilde N_1 Q + \tilde Q N_1) + \alpha_0\, \Phi_0 \Sigma_0 + \alpha_k\, \Phi_k \tilde N_2 (M_{22})^{k-1} N_2 \\
&\quad + \frac{1}{\hat\Lambda}\, [a_jb_iM_{ij} + a_i r \tilde N_i + \tilde r b_i N_i + \tilde r r \Sigma_0 ] \;.
\end{aligned} \end{equation}
This time we can integrate out $\Phi_0$ and $\Sigma_0$ as well. After doing it we obtain:
\begin{equation}
W' = \frac{1}{h\hat\Lambda^2}\, (a_1b_1a_2b_2 - a_1b_2a_2b_1) + \frac{1}{\hat\Lambda}\, (\tilde N_2 a_2 r + \tilde r b_2 N_2) + \frac{\alpha_k}{h^{k-1} \hat\Lambda^{k-1}}\, \Phi_k \tilde N_2 (a_1b_1)^{k-1} N_2 \;.
\end{equation}

The operators $\calO_k = \Phi_k \, \tilde q (A_2B_2)^k q$ behave quite differently from the previous $\tilde \calO_k$. They are still irrelevant, but actually dangerous irrelevant (see \cite{Strassler:2005qs} for a similar discussion in SQCD with quartic superpotential). Their quantum dimension becomes smaller and smaller going towards the IR, until some point when they behave as mass terms and the corresponding gauge singlet $\Phi_0$ is integrated out together with the would-be-generated gauge singlet $\Sigma_0$ in the opposite $(N_f, \overline{N_f})$ representation.

We can define a relative number $N_\Phi$ counting the number of $(\overline{N_f},N_f)$ fields ($\Phi_k$) minus the number of $(N_f, \overline{N_f})$ fields ($\tilde \Phi_k$). This number decreases by one unit at each Seiberg duality, either because a field contributing $+1$ is integrate out or because one contributing $-1$ is generated. We could say that our theory is self-similar along the cascade just adding this number $N_\Phi$ to the list of running ones.

\

The flow of the theory drastically depends on the choice of initial ranks, as also observed in \cite{Ouyang:2003df} (see also \cite{Benini:2007gx}). Since along the flow both ranks $g_1$ and $g_2$ and their difference reduce, we could either reach a point were one of the ranks is zero (or order of their difference), or a point where the difference is zero (or order $N_f$) while the ranks are still large. I am interested in the latter situation. Notice from \eqref{beta functions} that if $(g_1-g_2)<N_f/4$ both $\beta$-functions are positive and there are no Seiberg dualities anymore.

Thus we can imagine the following flow, from the bottom up. In the far IR the two gauge ranks are equal (say $N_0$), the theory has no exotic gauge singlet fields ($N_\Phi=0$) and there are no flavor-flavor-flavor (f-f-f) anomalies at all.
Both $\beta$-functions are positive. This theory was extensively studied in \cite{Benini:2006hh} where a proposal was made for the full flow down to a conformal fixed point with flavors. To go up in energy, we perform Seiberg dualities.%
\footnote{Recall that the RG flow is irreversible: the UV determines the IR but not the opposite. So we always have to think in terms of describing the theory at some scale, compatible with its flow to the IR.}
So at step one the gauge group is $SU(N_0 + N_f)\times SU(N_0)$ and there is one gauge singlet $\Phi_0$ (thus $N_\Phi=1$) with superpotential coupling $\calO_0$. Still there are no f-f-f anomalies. This theory correctly flows in the IR to what I stated above. Going generically up by $n$ steps, the gauge ranks are as prescripted by the cascade, and there are $n$ gauge singlets $\Phi_{k=0\dots n-1}$ ($N_\Phi=n$) with their corresponding superpotential couplings $\calO_k$.

In order to study this theory at strong coupling and for a large number of flavors ($N_f$ of order $N_c$), I am going to construct a supergravity dual to this flow.

\section{SUSY D7 Probes on the Warped Conifold}

My aim is now to realize a supergravity dual of the previous theory and its RG flow. The starting point is obviously the easiest of its steps, namely the $SU(g_1) \times SU(g_2)$ theory without extra gauge singlets. We are going to realize it as the near horizon theory of a stack of (fractional) D3-branes at a Calabi-Yau singularity plus non-compact D7-branes. Let me proceed stepwise.

As is well known, putting $N$ D3-branes at a conifold singularity \cite{Klebanov:1998hh} we realize an $SU(N)\times SU(N)$ \Nugual{1} gauge theory, with chiral fields $A_i$, $B_i$ ($i=1,2$) in the bifundamental and anti-bifundamental representation of the gauge group. On the gravity side, at large $N$ and in the near horizon limit the system is described by Type IIB Supergravity on a warp product space of 4d Minkowski and a 6d conifold, with warp factor $h(r) = L^4/r^4$ and $N$ units of self-dual 5-form flux.

As suggested in \cite{Karch:2002sh} and then further on investigated in \cite{Kruczenski:2003be}, we can add chiral superfields transforming in the fundamental representation of the gauge group by wrapping D7-branes on 4-cycles on the gravity side. In order that the gauge theory living on the D7-brane worldvolume describes a non-dynamical global flavor symmetry on the field theory side, the 4-cycle must be non-compact. And in order to have a supersymmetric embedding the 4-cycle must be holomorphic (I will have to say more about this). We can specify the complex structure of the conifold by defining it as the variety $z_1z_2 - z_3z_4 = 0$ in $\bbC^4$. Among the many, there are two classes of holomorphic divisors which are interesting for us.

The first class of 4-cycles is represented by $\Sigma_K = \{ z_1 + z_2 = 0 \}$. It preserves a diagonal $SU(2)_D$ subgroup and the $U(1)_R$ factor of the conifold isometry group $SU(2)^2\times U(1)_R$, and was studied in \cite{Kuperstein:2004hy}. This embedding was then used in \cite{Benini:2007gx} to add non-chiral matter to the KS \cite{Klebanov:2000hb} and KT \cite{Klebanov:2000nc} theories. The other class is represented by $\Sigma_O = \{ z_1 = 0 \}$ and was extensively studied in \cite{Ouyang:2003df}. The latter 4-cycle has very different properties from the former: it fully breaks the $SU(2)^2$ conifold isometry (but still preserving $U(1)_R$) and it is made of two separate intersecting branches. As argued in \cite{Ouyang:2003df} it introduces chiral matter exactly in the way we are looking for: according to the quiver of Figure \ref{fig:electric theory1} and with the superpotential \eqref{superpotential Ouyang}.

In order to create a disbalance in the gauge ranks we have to add D5-branes wrapped on the non-trivial 2-cycle of the conifold \cite{Gubser:1998fp}. Their presence generates, among the other effects, background values for the 3-form fluxes $F_3$ and $H_3$. The main difference between the two classes of D7-brane embeddings is that on the non-chiral $\Sigma_K$ the pull-back of $H_3$ is zero, whilst on the chiral $\Sigma_O$ is not. If $\hat H_3$ (hatted quantities are pulled-back) is zero we can always gauge away a possible pull-back of $B_2$ by a choice of $F_2$, so that $\calF=0$.%
\footnote{Some care has to be paid to possible sources for $\calF$ on the brane, arising whenever $\hat C_6 \neq 0$. Moreover $F_2$ is quantized on 2-cycles.}
I defined the gauge invariant flux on the brane as $\calF = \hat B_2 + 2\pi F_2$, where $F_2 = dA$ is the usual field strength of the gauge bundle. If $\hat H_3 \neq 0$ we cannot gauge away $\calF$ in any way and we have to worry about it. As we will see its effects are many: first of all it affects the supersymmetry constraints on the brane configuration, moreover it generates new induced charges and modifies the running of bulk fluxes.

\

The first step in the construction of a fully backreacted solution with this kind of D7-branes is to understand which are the supersymmetric embeddings and what is the flux induced. These two issues are addressed by studying probe branes.

To set our conventions, we start considering a probe D7-brane along $\Sigma_O = \{z_1=0\}$ in the singular conifold with 5-form flux (the KW theory \cite{Klebanov:1998hh}), and look for possible SUSY gauge bundles. The metric of the supergravity solution is
\begin{equation} \begin{aligned} \label{sing conif metric}
ds^2 &= h(r)^{-\frac{1}{2}} \, dx_{3,1}^2 + h(r)^{\frac{1}{2}} \Bigl\{ dr^2 + r^2 \, ds_{T^{1,1}}^2 \Bigr\} \\
ds_{T^{1,1}}^2 &= \frac{1}{6} \sum\nolimits_j \Bigl( d\theta_j^2 + \sin^2\theta_j \, d\varphi_j^2 \Bigr) + \frac{1}{9} ( d\psi + {\textstyle \sum_j \cos\theta_j \, d\varphi_j} )^2
\end{aligned} \end{equation}
where the warp factor is given by $h(r) = L^4/r^4$, with $L^4 = \frac{27}{16}4\pi g_sN \alpha'^2$. In the following we will set $\alpha'=1$. The Calabi-Yau (CY) geometry is described by a real $(1,1)$ K\"ahler form and an holomorphic $(3,0)$-form, both closed and co-closed:
\begin{align}
J &= \frac{1}{3} r\, dr\wedge (d\psi +  {\textstyle \sum_j \cos\theta_j\, d\varphi_j}) - \frac{1}{6} r^2 \Bigl[ \sin\theta_1 \, d\theta_1\wedge d\varphi_1 + \sin\theta_2 \, d\theta_2\wedge d\varphi_2 \Bigr] \label{Kahler form} \\
\Omega &= e^{i\psi} (dr + i\, r\frac{1}{3} g^5) \wedge r^2 \frac{1}{6} (d\theta_1 - i \sin\theta_1\, d\varphi_1) \wedge (d\theta_2 - i \sin\theta_2\, d\varphi_2) \label{holomorphic form} \;.
\end{align}
They satisfy $J^3 = \frac{3i}{4} \Omega\wedge \overline{\Omega}=6 \text{vol}_6$. More details on my conventions and the CY geometry are written in Appendix \ref{sec:conventions}.

As shown in \cite{Marino:1999af} the conditions for a spacetime filling D7-brane to be supersymmetric (which means that there is a $\kappa$-symmetry on the brane that preserves some Killing spinors of the bulk) on a background with closed NS-NS potential $B_2$ can be rephrased as:
\begin{itemize}
\item the embedding is holomorphic;
\item the gauge invariant field strength $\calF = \hat B_2 + 2\pi\alpha' F_2$ is a $(1,1)$-form;
\item it holds
\begin{equation} \label{kappa symmetry cond}
\hat J \wedge \calF = \tan\theta \, (\text{vol}_4 - \frac{1}{2} \calF\wedge\calF)
\end{equation}
for some constant $\theta$ (that depends on which combination of Killing spinors is preserved).%
\footnote{The expression for $\theta$ depends on how the 10d Killing spinors are constructed from the 6d one. In our class of $SU(3)$-structure solutions $\theta$ is a constant, but in general it can be a function of the 6d manifold. See an example in \cite{Benna:2006ib}.}
Here $J$ is the 6d K\"ahler form and $\text{vol}_4 = \frac{1}{2} \hat J\wedge\hat J$.
\end{itemize}
Then in \cite{Gomis:2005wc} it was shown that these conditions still assure $\kappa$-symmetry on a background $\calM_6$ with $SU(3)$-structure and NS-NS and R-R fluxes, provided that we substitute, in place of $J$, the 2-form $J_w$ that defines the $SU(3)$-structure of $\calM_6$. When $\calM_6$ is a warped CY, as in (\ref{sing conif metric}), $J_w = h^{1/2}\, J$.
% use the warped K\"ahler form $J_w \equiv h^{1/2}\, J$.

The holomorphic embedding we are considering is made of two branches: $\Sigma_j=\{ \theta_j, \varphi_j = \text{const} \}$. For definiteness we concentrate on $\Sigma_2$; then the pull-back of $J$ is easily derived from \eqref{Kahler form}. We are looking for gauge bundles on the D7-brane such that $\calF$ is a real $(1,1)$-form, closed and co-closed. We take the $(1,1)$ ansatz
\begin{equation} \label{1,1 ansatz}
\calF = f_1(r)\, \frac{1}{3} r\, dr\wedge \hat g^5 + f_2(r)\, \frac{1}{6} r^2 \sin\theta_1\, d\theta_1\wedge d\varphi_1 \;.
\end{equation}
When imposing closure and co-closure (it is a linear system) we get two solutions:
\begin{align}
\calF^{ASD} &= \frac{1}{3r^3} dr\wedge \hat g^5 + \frac{1}{6r^2} \sin\theta_1\, d\theta_1\wedge d\varphi_1 \label{ASD KW probe bundle}\\
\calF^{SD} &= \frac{1}{3} r\, dr\wedge \hat g^5 - \frac{1}{6} r^2 \sin\theta_1\, d\theta_1\wedge d\varphi_1 \;.
\end{align}
The first solution solves the $\kappa$-symmetry condition \eqref{kappa symmetry cond}:
it is anti-self-dual (ASD: $\calF = -\ast_4 \calF$) and primitive ($\calF \wedge \hat J = \calF \wedge \hat J_w =0$). The second one instead is not supersymmetric: it is self-dual (SD) and in fact proportional to the unwarped K\"ahler form ($\calF \propto \hat J = h^{-1/2}\, \hat J_w$) so that it cannot solve \eqref{kappa symmetry cond} unless the warp factor is constant.%
\footnote{This is consistent with the fact that on D3-brane backgrounds a self-dual bundle cannot be supersymmetric as it carries anti-D3 charge. Without D3-branes, instead, the warp factor is constant.}

\

Then we consider a probe D7-brane in the Klebanov-Tseytlin background \cite{Klebanov:2000nc}. This supergravity solution describes $N$ D3-branes and $M$ fractional D3-branes at the tip of a singular conifold. The fractional D3's can be thought as D5-branes wrapped on the 2-cycle of the conifold and shrunk at the origin. The dual gauge theory has the same quiver diagram as in the Klebanov-Witten case, but with gauge group $SU(N+M)\times SU(N)$. Thus the effect of the wrapped D5's is to increase only one rank.

The metric is still that of a warped singular conifold \eqref{sing conif metric}, but with different warp factor:
\begin{equation}
h(r) = \frac{27\pi\alpha'^2}{4r^4} \Bigl[ g_s N + \frac{3 (g_sM)^2}{2\pi} \, \Big( \frac{1}{4} + \log \frac{r}{r_0} \Big) \Bigr] \;.
\end{equation}
As well known, the logarithmic behavior is dual to the cascade of gauge ranks \cite{Klebanov:2000hb}. The fluxes are:
\begin{equation}
B_2 = \frac{3}{2} g_s M \, \omega_2 \log\frac{r}{r_0} \qquad\qquad H_3 = \frac{3}{2} g_s M \, \frac{1}{r} dr\wedge \omega_2 \qquad\qquad F_3 = \frac{M}{2} \omega_3 \;,
\end{equation}
where we define
\begin{equation} \label{def forms} \begin{aligned}
\omega_2 &= \frac{1}{2} (\sin\theta_1 \, d\theta_1\wedge d\varphi_1 - \sin\theta_2 \, d\theta_2\wedge d\varphi_2) \\
\omega_3 &= g^5\wedge\omega_2 = \frac{1}{2} (d\psi +  {\textstyle \sum_j \cos\theta_j\, d\varphi_j}) \wedge (\sin\theta_1 \, d\theta_1\wedge d\varphi_1 - \sin\theta_2 \, d\theta_2\wedge d\varphi_2) \\
\omega_5 &= -2\, \omega_2\wedge\omega_2 \wedge g^5 = \sin\theta_1\, d\theta_1\wedge d\varphi_1 \wedge \sin\theta_2 \, d\theta_2\wedge d\varphi_2 \wedge g^5 \\
g^5 &= (d\psi +  {\textstyle \sum_j \cos\theta_j\, d\varphi_j}) \;.
\end{aligned} \end{equation}
The 3-form fluxes are such that $g_s \ast_6 F_3 = H_3$.

In this case the pull-back of $H_3$ on the D7-brane is non-zero, and since $d\calF=\hat H_3$ we are forced to consider a non-trivial gauge flux. Thus I will use again for $\calF$ the $(1,1)$ ansatz of \eqref{1,1 ansatz} and impose that
\begin{equation}
d\calF = \hat H_3 \qquad\qquad \hat J \wedge \calF = 0 \;.
\end{equation}
The solution is
\begin{equation} \label{ASD KT probe bundle}
\calF^{ASD} = \Bigl( \frac{9g_sM}{4} \, \frac{1}{r^2} + \frac{C_1}{r^4} \Bigr) \Bigl( \frac{1}{3}r\,dr\wedge \hat g^5 + \frac{1}{6} r^2 \sin\theta_1\, d\theta_1\wedge d\varphi_1 \Bigr) \;,
\end{equation}
with $C_1$ an arbitrary constant. This flux is anti-self-dual. Notice that the homogeneous solution is exactly $\calF^{ASD}$ of \eqref{ASD KW probe bundle}.

Also in this case we could find a self-dual gauge flux with still $d\calF = \hat H_3$:
\begin{equation}
\calF^{SD} = \Bigl( \frac{9g_sM}{4} \, \frac{1}{r^2} + C_2 \Bigr) \Bigl( \frac{1}{3}r\,dr\wedge \hat g^5 - \frac{1}{6} r^2 \sin\theta_1\, d\theta_1\wedge d\varphi_1 \Bigr) \;.
\end{equation}
Again this configuration is not supersymmetric.

\section{Type IIB Supergravity with Sources}

In the previous Section we understood that the Klebanov-Tseytlin background supports probe D7-branes which are spacetime-filling, non-compact, supersymmetric and along the embeddings we need to realize the chiral cascading field theory we are interested in. Such branes, in order to be SUSY, need to have a non-trivial anti-self-dual gauge flux $\calF$ on them. We are going to construct a fully backreacted solution for this system; thus I report how the Type IIB Supergravity equations of motion (EOM) are modified in presence of these sources.

The action of IIB Supergravity with D7-branes in Einstein frame is in my conventions:
\begin{equation} \begin{aligned}
S_{IIB} &= \frac{1}{2\kappa_{10}^2} \int d^{10}x \, \sqrt{-g} \Bigl\{ R - \frac{1}{2} |\partial\phi|^2 -\frac{1}{2} e^{-\phi} |H_3|^2 -\frac{1}{2} e^{2\phi} |F_1|^2 - \frac{1}{2} e^{\phi} |F_3|^2 -\frac{1}{4} |F_5|^2 \Bigr\} \\
&\quad -\frac{1}{4\kappa_{10}^2} \int C_4 \wedge H_3 \wedge F_3 \\
&\quad - \mu_7 \int_{D7} d^8\xi \, e^\phi \, \sqrt{-\det(\hat g+ e^{-\phi/2} \, \calF)} \; + \mu_7 \int \hat C_q \, e^{-\calF}
\end{aligned} \end{equation}
where the gauge invariant R-R field strengths are $F_p = dC_{p-1} - C_{p-3}\wedge H_3$.%
\footnote{The minus sign in $e^{-\calF}$ is related to the sign in the definition of $F_p$ and is required in order to obtain EOM's consistent with $d^2=0$.}
Moreover $2\kappa_{10}^2 = (2\pi)^7 \alpha'^4$, $\mu_p = (2\pi)^{-p} \alpha'^{-\frac{p+1}{2}}$ is the Dp-brane charge and tension so that $2\kappa_{10}^2 \mu_p = (4\pi^2\alpha')^\frac{7-p}{2}$. In particular $2\kappa_{10}^2\mu_7=1$. Hatted quantities are pulled-back.

Without sources (the last line) the Bianchi identities (BI) and EOM's for the form-fields are readily derived:
\begin{equation} \begin{aligned} \label{modified BI EOM}
dF_1 &= 0 \\
dF_3 &= H_3 \wedge F_1 \\
dF_5 &= H_3 \wedge F_3 \\
dH_3 &= 0
\end{aligned} \qquad\qquad\qquad \begin{aligned}
d\big( e^{2\phi}\ast F_1 \big) &= - e^\phi \, H_3\wedge \ast F_3 \\
d\big(e^\phi \ast F_3 \big) &= - H_3\wedge F_5 \\
d\ast F_5 &= dF_5 = H_3\wedge F_3 \\
d\big( e^{-\phi}\ast H_3 \big) &= e^\phi \, F_1\wedge\ast F_3 - F_5\wedge F_3 \;.
\end{aligned} \end{equation}
These have to be supplemented with the equation $F_5 = \ast F_5$, which is not derived from the action (see \cite{Belov:2006jd} for a solution to this problem). Notice that these BI's and EOM's are consistent with $d^2=0$. By comparing the EOM's with the BI's of the dual field strengths we get the relations
\begin{equation}
F_7 = - e^\phi \ast F_3 \qquad\qquad\qquad F_9 = e^{2\phi} \ast F_1 \;.
\end{equation}

Then we consider the effect of sources. The details of the derivation of the following Bianchi identities and equations of motion, as long as other formulae useful in computations, can be found in Appendix \ref{sec:EOM}. I find:
\begin{equation} \begin{aligned}
dF_1 &= -\Omega_2 &
d\big( e^{2\phi}\ast F_1 \big) &= - e^\phi \, H_3\wedge \ast F_3 - \frac{1}{24} \calF^4\wedge\Omega_2 \\
dF_3 &= H_3 \wedge F_1 -\calF\wedge\Omega_2 &
d\big(e^\phi \ast F_3 \big) &= - H_3\wedge F_5 + \frac{1}{6} \calF^3\wedge\Omega_2 \\
dF_5 &= H_3 \wedge F_3 - \frac{1}{2} \calF\wedge\calF\wedge\Omega_2 \qquad &
d\ast F_5 &= dF_5 \\
dH_3 &= 0 &
d\big( e^{-\phi}\ast H_3 \big) &= e^\phi \, F_1\wedge\ast F_3 - F_5\wedge F_3 + (\text{sources}) \;.
\end{aligned} \end{equation}
The 2-form $\Omega_2$ is a localized form orthogonal to the D7-brane such that
\begin{equation}
\int_{D7} X_8 = \int_{M_{10}} X_8 \wedge \Omega_2
\end{equation}
for every 8-form $X_8$ on the D7-brane. For an holomorphic embedding $\calC = \{f(z_j)=0\}$ the form can be written as $\Omega_2 = -i\, \delta^2 (f,\bar f) \, df\wedge d\bar f$. Thus in general for localized (and even smeared) holomorphic D7-branes $\Omega_2$ is a closed real $(1,1)$-form. Moreover it \emph{must} be exact in order to solve the BI of $F_1$, and this condition is precisely tadpole cancellation. The equation of motion of $H_3$ gets many contributions from the D7's and the complete expression is in (\ref{eom H3}). Notice again that the full system of equations is consistent with $d^2 = 0$.

In Appendix \ref{sec:EOM} the reader can also find a proof that the $\kappa$-symmetry condition \eqref{kappa symmetry cond} together with supersymmetry in the bulk assure that the EOM for the gauge connection on the D7-brane is satisfied. This was also shown on more general ground in \cite{Koerber:2005qi}. On the other hand in \cite{Koerber:2007hd} it was shown that supersymmetry and Bianchi identities implies the satisfaction of the EOM's for the form-fields, for the dilaton and of Einstein equation, for localized as well as smeared backreacting branes.

\section{The Backreacted Solution}

We have now collected enough elements to write down the backreacted solution. From the probe analysis we learned that the D7-branes source D7-charge as well as D5- and D3-charge, due to the non-trivial gauge flux $\calF$ on them. The gauge invariant flux $\calF$ is constrained to be $(1,1)$ and primitive. Then we only have to produce an ansatz and set to zero the supersymmetry variations in the bulk, as well as imposing BI's and EOM's for the form-fields.

As observed in many previous works of this kind \cite{Casero:2006pt, Casero:2007jj, Paredes:2006wb, Benini:2006hh, Benini:2007gx}, finding the fully backreacted solution for a system with color and flavor branes on a topologically non-trivial manifold is a very challenging task, due to the low amount of symmetry. In general, and in our case too, the addition of non-compact D7-branes breaks some symmetries of the background where they are put; consequently one should write a complicated ansatz which would lead to partial differential equations, difficult or impossible to solve. My main tool will be an angular smearing.

The procedure is the same as in \cite{Casero:2006pt, Casero:2007jj, Paredes:2006wb, Benini:2006hh, Benini:2007gx}. Our D7-branes are put along $\Sigma_1 = \{\theta_1,\varphi_1 = \text{const}\}$ and $\Sigma_2 = \{\theta_2,\varphi_2 = \text{const}\}$: each branch is localized at a point of one of the $S^2$ over which the $T^{1,1}$ fibration is constructed. A single D7-brane breaks the $SU(2)^2\times U(1)$ isometry of the conifold to $U(1)^3$.
Since we are going to put a large number of flavor branes and the preserved Killing spinors are independent from the particular point chosen, we put each brane at a different location restoring the original $SU(2)^2\times U(1)$ isometry. The ansatz we propose will thus have this same isometry group. The smeared charge distribution for $N_f$ D7-branes, each made of two branches, is (\cite{Benini:2006hh})
\begin{equation}
\Omega_2 = \frac{N_f}{4\pi} (\sin\theta_1\, d\theta_1\wedge d\varphi_1 + \sin\theta_2\, d\theta_2\wedge d\varphi_2) \;.
\end{equation}

The metric ansatz is
\begin{equation} \begin{aligned} \label{metric ansatz}
ds^2 &= h(\rho)^{-\frac{1}{2}} \, dx_{3,1}^2 + h(\rho)^{\frac{1}{2}} \, ds_6^2 \\
ds_6^2 &= e^{2u(\rho)} \Bigl( d\rho^2 + \frac{1}{9} ( d\psi + {\textstyle \sum_j \cos\theta_j \, d\varphi_j} )^2 \Bigr) + e^{2g(\rho)} \frac{1}{6} \sum\nolimits_j \Bigl( d\theta_j^2 + \sin^2\theta_j \, d\varphi_j^2 \Bigr)
\end{aligned} \end{equation}
which depends on three unknown functions $u(\rho)$, $g(\rho)$ and $h(\rho)$. Led by the Bianchi identity $dF_1 = -\Omega_2$ we put
\begin{equation}
F_1 = \frac{N_f}{4\pi} g^5 \;.
\end{equation}
The ansatz for $B_2$ is as in the KT solution, because D7-branes do not source any F1-charge:
\begin{align}
B_2 &= \Bigl(\frac{M}{2} f(\rho)+\pi\, b_2^{(0)} \Bigr) \, \omega_2 \\
H_3 &= \frac{M}{2} f'(\rho) \, d\rho\wedge \omega_2 \;.
\end{align}
We put a constant shift in $B_2$ for later convenience. Our solution will have $\lim_{\rho\to -\infty} f(\rho)=0$, so that $b_2^{(0)}$ represents the constant value in the far IR. We will see in Section \ref{sec:the cascade} which is the meaning of the constant $M$.

In order to compute the gauge flux on a \emph{single} D7-brane we need the 6d unwarped K\"ahler form:
\begin{equation}
J_6 = \frac{1}{3} e^{2u} d\rho \wedge g^5 - \frac{1}{6} e^{2g} (\sin\theta_1\, d\theta_1\wedge d\varphi_1 + \sin\theta_2\, d\theta_2\wedge d\varphi_2) \;.
\end{equation}
which is directly derived from the metric. Then we can write the gauge flux $\calF$ on each brane. It must satisfy $d\calF = \hat H_3$ and, in order to preserve $\kappa$-symmetry, it must be real $(1,1)$ and primitive ($\calF\wedge \hat J=0$). Let me start considering the branch $\Sigma_2$. The $\kappa$-symmetry constraints are easily encoded in the ansatz
\begin{equation} \label{fieldstrength 2 probe}
\calF \Bigr|_{\Sigma_2} = p(\rho) \Bigl[ \frac{1}{3} e^{2u} d\rho \wedge \hat g^5 + \frac{1}{6} e^{2g} \, \sin\theta_1\, d\theta_1\wedge d\varphi_1 \Bigr] \;,
\end{equation}
which is also consistent with the $SU(2)\times SU(2)$ symmetry of the field theory.
Then the relation $d\calF = \hat H_3$ gives the following equation:
\begin{equation} \label{fieldstrength eq1}
\frac{M}{4} f'(\rho) = \frac{1}{3} e^{2u(\rho)} p(\rho) + \frac{1}{6} \, \frac{\partial}{\partial\rho} \Bigl( e^{2g(\rho)} p(\rho) \Bigr) \;.
\end{equation}
On the other branch $\Sigma_1$ the gauge flux is the same but with opposite sign, namely:
\begin{equation}
\calF \Bigr|_{\Sigma_1} = - p(\rho) \Bigl[ \frac{1}{3} e^{2u} d\rho \wedge \hat g^5 + \frac{1}{6} e^{2g} \, \sin\theta_2\, d\theta_2\wedge d\varphi_2 \Bigr] \\
\end{equation}
with the same function $p(\rho)$ as before.

I conclude the ansatz with an expression of $F_3$ which automatically solves its Bianchi identity $dF_3 = H_3\wedge F_1 - \calF\wedge \Omega_2$. Here we have to put some care in the computation of the effect of the smearing on $\calF\wedge\Omega_2$, starting from the localized expressions for the two branches. For each branch, the localized charge distribution is a sum of delta functions at the different locations of the $N_f$ branes on the sphere: \label{discussion smear}
%\begin{equation}
$ \Omega_2^{loc} = \sum_{a=1}^{N_f} \delta^{(2)}( \theta_j - \theta_j^{(a)}, \varphi_j - \varphi_j^{(a)}) \, d\theta_j\wedge d\varphi_j $.
%\end{equation}
Here $\theta_j^{(a)}$ and $\varphi_j^{(a)}$ are the coordinates of the $a$-th brane, branch $j$. In the smearing we substitute such sum of delta functions with the homogeneous distribution $\Omega_2^{smeared} = \frac{N_f}{4\pi} \sin\theta_j\, d\theta_j\wedge d\varphi_j$. We simply have to repeat the same procedure for $\calF\wedge\Omega_2$:
\begin{equation}
\calF^{(\Sigma_j)} \wedge \Omega_2^{(\Sigma_j)} = \calF^{(\Sigma_j)} \wedge \delta^{(2)}(\theta_j , \phi_j) \, d\theta_j\wedge d\varphi_j \;\to\; \calF^{(\Sigma_j)} \wedge \frac{N_f}{4\pi} \sin\theta_j\, d\theta_j\wedge d\varphi_j
\end{equation}
Summing the contributions from the two branches, we eventually get:
\begin{equation}
(\calF \wedge \Omega_2)^\text{smeared} = - \frac{N_f}{6\pi} e^{2u} p(\rho) \, d\rho\wedge g^5 \wedge \omega_2 \;.
\end{equation}

Here it is worth stressing a subtle point. Naively one could have thought that since $H_3 \wedge \Omega_2^\text{smeared} = 0$ then there is no pull-back of $H_3$ on the smeared configuration of branes, and thus it is consistent to put their gauge flux to zero. But, as we saw, this is not actually correct. What is correct is computing the flux on a single (probe) brane, then evaluate $\calF \wedge \Omega_2^\text{loc}$ and smear the latter. The content of $H_3\wedge \Omega_2^\text{smeared}=0$ is that, in fact, $d(\calF\wedge\Omega_2)^{smeared} = H_3 \wedge \Omega_2^\text{smeared} = 0$.

Eventually, using equation \eqref{fieldstrength eq1} we obtain
\begin{equation}
H_3\wedge F_1 - (\calF\wedge\Omega_2)^\text{smeared} = \frac{M N_f}{8\pi} \, \parfrac{}{\rho} \Bigl[ f + f - \frac{2}{3M} e^{2g} p \Bigr] \, d\rho\wedge g^5 \wedge \omega_2 \;.
\end{equation}
It is nice to observe that $H_3\wedge F_1$ contributes $f$ in brackets while $(\calF\wedge\Omega_2)^\text{smeared}$ contributes the other $f$. This doubling with respect to the non-chiral case discussed in \cite{Benini:2007gx} (where the term $(\calF\wedge\Omega_2)^\text{smeared}$ is not present) is dual in field theory to the fact that in the chiral theory the difference of gauge ranks gets reduced by $N_f$ at each step of the cascade, while in the non-chiral theory it scales by $N_f/2$.

The ansatz for $F_3$ is then
\begin{equation}
F_3 = \frac{M}{2} \Bigl[ \frac{N_f}{2\pi} f - \frac{N_f}{6\pi M} e^{2g} p \Bigr] \, g^5 \wedge \omega_2 \;.
\end{equation}
Notice that we should have allowed an integration constant $C$ in brackets; this constant can be absorbed in a redefinition of $f(\rho)$ and then appears in $B_2$, as we accordingly took into account.

For completeness I report the expression of the gauge field strength and connection on the branes, as derived from the definition $\calF = \hat B_2 + 2\pi F_2$ and equation \eqref{fieldstrength eq1}:
\begin{align}
2\pi F_2 \Bigr|_{\Sigma_2} &= \frac{1}{3} e^{2u} p \, d\rho\wedge\hat g^5 + \Bigl( \frac{1}{6} \, e^{2g}p - \frac{M}{4} f - \frac{\pi}{2} \, b_2^{(0)} \Bigr) \sin\theta_1\, d\theta_1\wedge d\varphi_1 \label{fieldstrength F2} \\
2\pi A \Bigr|_{\Sigma_2} &= \Bigl( \frac{M}{4} f - \frac{1}{6} \, e^{2g} p + \frac{\pi}{2} \, b_2^{(0)} \Bigr) \, \hat g^5 \;. \label{connection A}
\end{align}
The expressions on $\Sigma_1$ are the same but with opposite sign.

\

Now that the ansatz is complete we can solve it. We impose that the supersymmetry variations vanish. The details of the computation can be found in Appendix \ref{sec:SUSY variations}. We find that the equations for the 3-form flux decouple from the other ones, that can be solved first. Being the ansatz the same as in \cite{Benini:2006hh}, the equations and their solutions are also the same. We find the system
\begin{equation} \label{geometric system}
\left\{ \begin{aligned}
\phi' &= \frac{3N_f}{4\pi} e^\phi \\
g' &= e^{2u-2g} \\
u' &= 3 - 2e^{2u-2g} - \frac{3N_f}{8\pi} e^\phi
\end{aligned} \right.
\end{equation}
which can be (explicitly) integrated first. Its solution is%
\footnote{I suppress many integration constants. For a general discussion see \cite{Benini:2006hh}.}
\begin{equation}
e^\phi = \frac{4\pi}{3N_f}\, \frac{1}{(-\rho)}
\qquad\qquad\qquad \begin{aligned}
e^{2u} &= -6\rho (1-6\rho)^{-2/3} \, e^{2\rho} \\
e^{2g} &= (1-6\rho)^{1/3} \, e^{2\rho} \;.
\end{aligned} \end{equation}
The range of the radial coordinate is $\rho \in (-\infty,0]$; $\rho = -\infty$ corresponds to the IR while $\rho=0$ is an UV duality wall. The equations for the 3-form flux impose that the combination $G_3 \equiv F_3 - i\, e^{-\phi} H_3$ is imaginary-self-dual, that is $e^\phi \ast_6 F_3 = H_3$. Notice that it is also primitive by construction. We get
\begin{equation} \label{ISD eq2}
e^\phi \, \frac{3M}{4} \Bigl[ \frac{N_f}{2\pi} f - \frac{N_f}{6\pi M} \, e^{2g} p \Bigr] = \frac{M}{4} f' \;.
\end{equation}

The equations for the gauge flux \eqref{fieldstrength eq1} and the 3-form flux \eqref{ISD eq2} can be rewritten in terms of $\tilde p \equiv e^{2g}p$ and $\tilde f \equiv e^{-2\phi}f$. We write the second one and their difference:
\begin{equation}
\left\{ \begin{aligned}
&\tilde p = -\frac{2\pi M}{N_f}\, e^{\phi} \, \tilde f' \\
&\frac{2}{3M} \Bigl[ 2\, e^{2u-2g} \, \tilde p + \tilde p' \Bigr] = e^\phi \Bigl[ \frac{3N_f}{2\pi}\, e^{2\phi}\, \tilde f - \frac{N_f}{2\pi M}\, \tilde p \Bigr] \;.
\end{aligned} \right.
\end{equation}
% \begin{equation}
% \left\{ \begin{aligned}
% &\tilde p = -\frac{2\pi M}{N_f} e^{\phi} \, \tilde f' \\
% &\frac{2}{3M} \Bigl[ 2 e^{2u-2g} \, \tilde p - \tilde p' \Bigr] = e^\phi \Bigl[ \frac{6N_f}{4\pi} e^{2\phi} \tilde f + \frac{N_f}{2\pi M} \tilde p \Bigr] \;.
% \end{aligned} \right.
% \end{equation}
Substituting the first into the second we get a second order linear ODE:
\begin{equation} \label{diff eqn}
\tilde f'' + 2\Bigl( \frac{3N_f}{4\pi} e^\phi + e^{2u-2g} \Bigr) \, \tilde f' + 2 \Bigl( \frac{3N_f}{4\pi} \Bigr)^2 e^{2\phi} \, \tilde f = 0 \;,
\end{equation}
where we could also substitute the actual profile of the functions. The equation can be analytically integrated. Let me remark the dependence of the functions on $M$ and $N_f$ before: if we take $f$ of order one then $\tilde f$ is of order $N_f^2$ and $\tilde p$ is of order $M$.

The author of \cite{Ouyang:2003df} tackles the same problem as here: the addition of D7-branes to the Klebanov-Tseytlin background. He computes the effect of the branes at leading order in $N_f/M$ as a perturbation of the original background. Since his procedure is quite different from mine, I comment on this in Appendix \ref{sec:comparison Ouyang}.

\subsection{Solutions}

Equation \eqref{diff eqn} is a second order linear ODE, so there is a two dimensional vector space of solutions. The first solution is
\begin{equation} \begin{aligned}
f &= \frac{(1-6\rho)^{2/3}}{-\rho} \, e^{-2\rho} \\
\tilde p &= \frac{3M}{2} \, \frac{12\rho^2-12\rho+1}{(-\rho)(1-6\rho)^{1/3}} \, e^{-2\rho}
\end{aligned} \qquad\qquad \begin{aligned}
p &= \frac{3M}{2} \, \frac{12\rho^2-12\rho+1}{(-\rho)(1-6\rho)^{2/3}} \, e^{-4\rho} \;.
\end{aligned} \end{equation}
% \tilde f &= \Bigl(\frac{3N_f}{4\pi}\Bigr)^2 (-\rho) (1-6\rho)^{2/3} e^{-2\rho} \\
% \tilde f' &= - \Bigl(\frac{3N_f}{4\pi}\Bigr)^2 \frac{12\rho^2-12\rho+1}{(1-6\rho)^{1/3}} e^{-2\rho}
Actually this is not the solution physically relevant for us, because both the 3-form flux and the gauge flux diverge in the IR (while we would like them to vanish, according to the field theory discussion). Nevertheless we can notice some interesting features. In the IR (large $|\rho|$) the function $f$ is suppressed by $1/(-\rho)$ with respect to $\tilde p/M$; thus the gauge bundle dominates over the 3-form flux and determines the IR physics. In fact using the approximate IR relation $\log\rho = r$ we get the ASD solution \eqref{ASD KW probe bundle} in the KW background.

The second solution is expressed  in terms of the $E_n(z)$ function%
\footnote{In Mathematica is called ExpIntegralE.}
defined as
\begin{equation}
E_n(z) = \int_1^\infty \frac{e^{-z\, t}}{t^n} dt = \int_0^1 e^{-\frac{z}{\eta}} \eta^{n-2} d\eta \;.
\end{equation}
For completeness here are some of its properties:
\begin{equation}
\partial_z E_n(z) = - E_{n-1}(z) \qquad\qquad\qquad
n\, E_{n+1}(z) = e^{-z} - z\, E_{n}(z)
\end{equation}
and the series expansions around $z\to0$ and $z\to\infty$:
\begin{equation} \begin{aligned}
z\to 0: \quad E_n(z) &= z^{n-1} \Gamma(1-n) + \sum_{j=0}^\infty \frac{(-1)^{j+1}}{j!(j+1-n)}z^j \\
z\to \infty: \quad E_n(z) &= \frac{e^{-z}}{z} \Bigl[ \sum_{j=0}^\infty (-1)^j \frac{\Gamma(n+j)}{\Gamma(n)} \frac{1}{z^j} \Bigr] = \frac{e^{-z}}{z} + \calO \Bigl( \frac{e^{-z}}{z^2} \Bigr) \;.
\end{aligned} \end{equation}
% where $\Gamma(n+j)/\Gamma(n)=n(n+1)\dots(n+j-1)$.

The solution is:
\begin{equation} \begin{aligned}
f &= \frac{1}{(-\rho)} \, \Bigl[ 3-(1-6\rho) \, e^{\frac{1}{3}-2\rho} \, E_{2/3} \Bigl( \frac{1}{3}-2\rho \Bigr) \Bigr] \\
\tilde p &= \frac{3M}{2} \, \frac{1}{(-\rho)} \, \Bigl[ 3-6\rho - (12\rho^2 - 12\rho +1) \, e^{\frac{1}{3}-2\rho} \, E_{2/3} \Bigl( \frac{1}{3}-2\rho \Bigr) \Bigr] \\
p &= \frac{3M}{2} \, \frac{e^{-2\rho}}{(-\rho)(1-6\rho)^{1/3}} \, \Bigl[ 3-6\rho - (12\rho^2 - 12\rho +1) \, e^{\frac{1}{3}-2\rho} \, E_{2/3} \Bigl( \frac{1}{3}-2\rho \Bigr) \Bigr] \;.
\end{aligned} \end{equation}
% \tilde f &= \Bigl(\frac{3N_f}{4\pi}\Bigr)^2 (-\rho) \Bigl[ 3-(1-6\rho)e^{\frac{1}{3}-2\rho} E_{2/3} \Bigl( \frac{1}{3}-2\rho \Bigr) \Bigr] \\
% \tilde f' &= -\Bigl(\frac{3N_f}{4\pi}\Bigr)^2 \Bigl[ 3-6\rho - (12\rho^2 - 12\rho +1) e^{\frac{1}{3}-2\rho} E_{2/3} \Bigl( \frac{1}{3}-2\rho \Bigr) \Bigr] \\

\begin{figure}[t]
\begin{center}
\hspace{\stretch{1}}
\includegraphics[width=.35\textwidth]{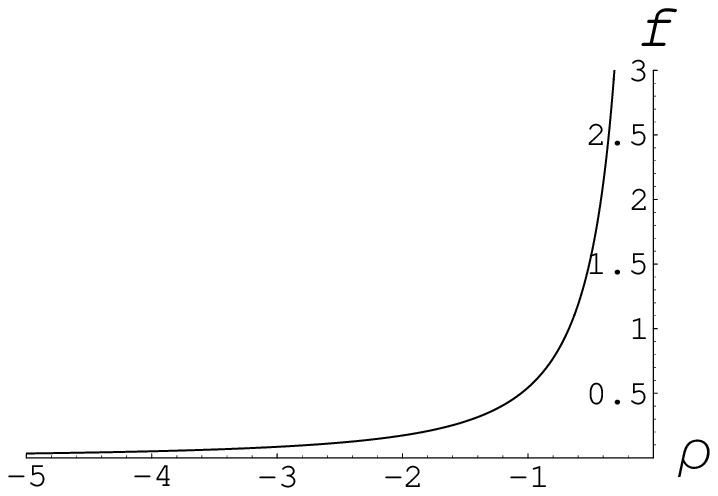}
\hspace{\stretch{1}}
\includegraphics[width=.35\textwidth]{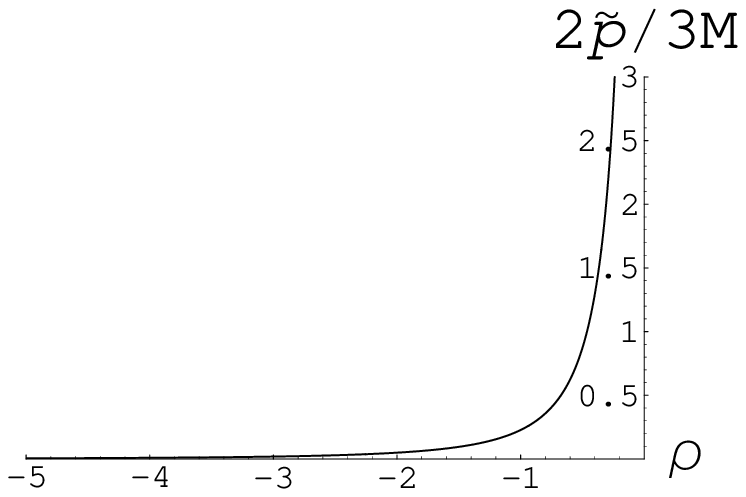}
\hspace{\stretch{1}}

\hspace{\stretch{1}}
\includegraphics[width=.35\textwidth]{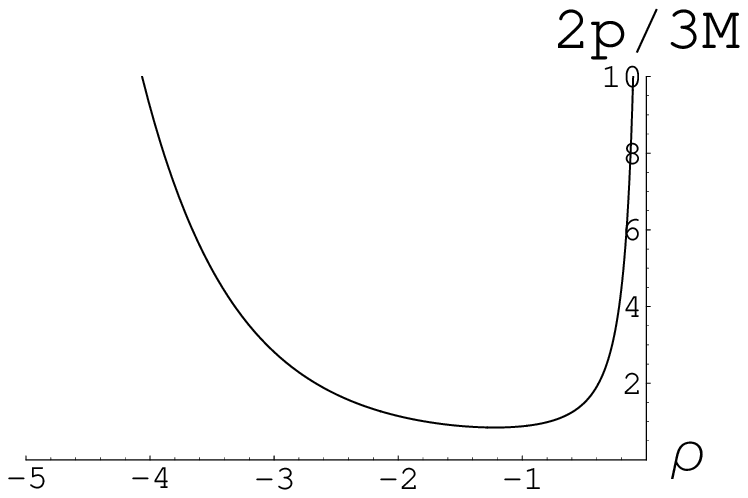}
\hspace{\stretch{1}}
\includegraphics[width=.35\textwidth]{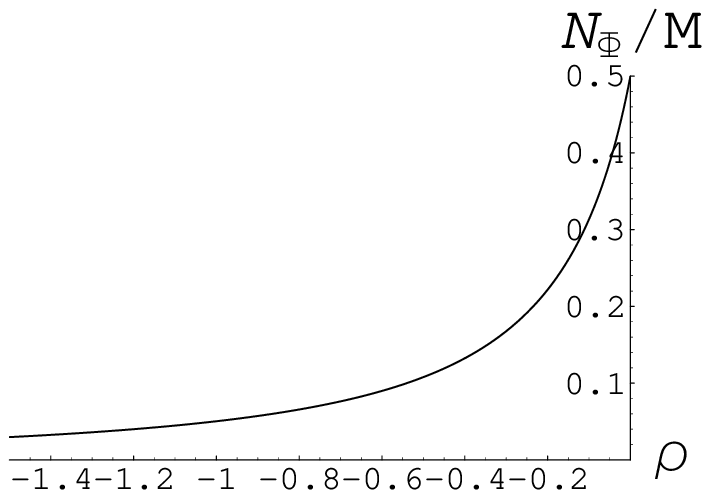}
\hspace{\stretch{1}}
\caption{Plot of some relevant functions: $f(\rho)$, $2\tilde p(\rho)/3M$, $2p(\rho)/3M$ and  $N_\Phi(\rho)$. \label{fig: functions}}
\end{center}
\end{figure}

The expansions of $f(\rho)$ and $\tilde p(\rho)$ around $\rho\to -\infty$ (IR) and $\rho\to 0^-$ (UV) are:
\begin{equation} \begin{aligned}
IR: \qquad &\begin{aligned}
f &= \frac{1}{\rho^2} + \frac{1}{\rho^3} + \frac{17}{12}\frac{1}{\rho^4} + \calO\Bigl( \frac{1}{\rho^5} \Bigr) \\
\tilde p &= \frac{3M}{2} \Bigl\{ -\frac{1}{\rho^3} - \frac{17}{6}\frac{1}{\rho^4} + \calO\Bigl( \frac{1}{\rho^5} \Bigr) \Bigr\}
\end{aligned} \\
UV: \qquad &\begin{aligned}
f &=  -\frac{\alpha}{\rho} -6(2-\alpha) -6\alpha\, \rho + \calO(\rho^2) \\
\tilde p &= \frac{3M}{2} \Bigl\{ -\frac{\alpha}{\rho} -12(2-\alpha) -18\alpha\, \rho + \calO(\rho^2) \Bigr\}
\end{aligned}
\end{aligned} \end{equation}
with $\alpha = 3-e^{1/3} E_{2/3}(1/3) \simeq 1.48$. The plots of all these functions are in Figure \ref{fig: functions}.

In this case, in the IR the function $\tilde p/M$ is negligible with respect to $f$, and the solution asymptotes the non-homogeneous piece of the ASD probe solution \eqref{ASD KT probe bundle} in the KT background.

\subsection{5-form Flux and Warp Factor}

The ansatz for the self-dual 5-form flux is related to the warp factor in the usual way. This is imposed by supersymmetry in the bulk, and we set:
\begin{equation} \label{F5 ansatz}
F_5 = (1+\ast)\, d^4x\wedge dh^{-1} = d^4x\wedge dh^{-1} + \frac{h' e^{4g}}{108} \, \omega_5
\end{equation}
with $\omega_5$ defined in \eqref{def forms}. We solve the Bianchi identity $dF_5 = H_3\wedge F_3 - \frac{1}{2} \calF\wedge\calF\wedge \Omega_2$. The first term is readily computed. In order to evaluate $(\calF \wedge \calF \wedge \Omega_2)^{smeared}$ we proceed as before: we first compute the localized expressions for the two branches and then we sum them, obtaining:
\begin{equation}
\big( -\tfrac{1}{2} \calF\wedge\calF\wedge\Omega_2 \big)^\text{smeared} = - \frac{N_f}{36\pi}\, p^2\, e^{2u+2g}\, d\rho\wedge \omega_5 \;.
\end{equation}
Combining the two pieces we get the second order equation:
\begin{equation}
\parfrac{}{\rho} \Bigl( \frac{h'\, e^{4g}}{108} \Bigr) = - \frac{M^2N_f}{16\pi} \, f' \Bigl( f-\frac{1}{3M}\tilde p \Bigr) - \frac{N_f}{36\pi}\, p^2\, e^{2u+2g} \;.
\end{equation}

As we expect from supersymmetry, this equation can be integrated to a first order equation. Making use of the BPS equation \eqref{fieldstrength eq1} we get:
\begin{equation}
\frac{h'\, e^{4g}}{108} = -\frac{\pi}{4}\, N_0 - \frac{M^2N_f}{32\pi}\, f^2 + \frac{N_f}{144\pi}\, \tilde p^2 \;.
\end{equation}
Here $N_0$ is an integration constant. This expression also fixes the effective D3-charge, and $N_0$ represents the D3-charge in the far IR.

\

The warp factor is obtained by integration. Thus:
\begin{equation}
h(\rho) = \int_\infty^\rho 108 \, e^{-4g(x)} \Bigl\{ -\frac{\pi}{4}\, N_0 - \frac{M^2N_f}{32\pi}\, \Bigl[ f(x)^2 - \frac{1}{2} \Bigl( \frac{2\tilde p(x)}{3M} \Bigr)^2 \Bigr] \Bigr\} dx
\end{equation}
As in \cite{Benini:2006hh,Benini:2007gx} the integration constant is chosen such that the analytic continuation of $h$ at plus infinity vanishes.
This expression cannot be analytically integrated, but we can provide the expansions in the IR and the UV. We find:
\begin{equation} \begin{aligned}
IR: \qquad h &= 27\pi N_0 \frac{e^{-4\rho}}{(-6\rho)^{2/3}} \Bigl[ 1 + \orders{\frac{1}{\rho}} \Bigr] \\
UV: \qquad h &= - \frac{27}{16\pi} M^2 N_f \frac{\alpha^2}{(-\rho)} \, [1+\calO(\rho)]
\end{aligned} \end{equation}
In the IR we recognize the almost conformal behavior of the flavored KW solution \cite{Benini:2006hh}. In the UV the warp factor diverges to negative values, signaling that at some $\rho<0$ it becomes zero and the supergravity description breaks down, as in the UV region of the KS and KT solutions flavored with non-chiral fundamental matter \cite{Benini:2007gx}.

From Figure \ref{fig: functions} and the plot of $p$, one could think that the worldvolume flux diverges in the IR, invalidating the solution. Instead what matters is the modulus $|\calF|^2$ computed with the full 10d metric, including the warp factor. One gets in the IR:
\be
IR: \qquad |\calF|^2 = 2\, \frac{p^2}{h} = \frac{M^2}{6\pi N_0}\, \frac{1}{(-\rho)^6} + \orders{\frac{1}{\rho^7}} \;.
\ee
Thus the flux vanishes and its energy is integrable in the IR.

\section{Charges in Supergravity}

We go on with the analysis of the solutions just found. My main goal in this section is to match the cascade and the running of gauge ranks between supergravity and field theory. The way I proceed is similar to the analysis performed in \cite{Benini:2007gx}. We start computing Maxwell charges, defined as the integral of the corresponding R-R fluxes:
\begin{equation} \begin{aligned} \label{def eff charges}
N_{D3} &= - \frac{1}{(4\pi^2\alpha')^2} \int_{\calC_5} F_5 \qquad\qquad &
N_{D7} &= - \int_{\calC_1} F_1 \\
N_{D5} &= \frac{1}{4\pi^2\alpha'} \int_{\calC_3} F_3 \;.
\end{aligned} \end{equation}
For sign conventions see Appendix \ref{sec:EOM}, and we again set $\alpha'=1$.

The integration is done on the 3-cycle $S^3 = \{\theta_2,\varphi_2= \text{const}\}$ in $T^{1,1}$ and on the whole $T^{1,1}$ respectively. We compute the integral of $B_2$ on the 2-cycle $S^2 = \{ \theta_1 = \theta_2, \varphi_1 = -\varphi_2 \}$ as well, obtaining:
\begin{equation} \begin{aligned}
M_{eff} &= \frac{1}{4\pi^2} \int_{S^3} F_3 = \frac{M N_f}{2\pi} \Bigl( f-\frac{1}{3M} \tilde p\Bigr) \\
N_{eff} &= -\frac{1}{16\pi^4} \int_{T^{1,1}} F_5 = N_0 + \frac{M^2N_f}{8\pi^2}\, f^2 - \frac{N_f}{36\pi^2}\, \tilde p^2 \\
b_0 &= \frac{1}{4\pi^2} \int_{S^2} B_2 = \frac{M}{2\pi}\, f + b_2^{(0)} \;.
\end{aligned} \end{equation}

The integral $b_0$ is an axionic field defined modulo 1. Shifting it by 1 does not affect physical quantities; nonetheless it corresponds to a Seiberg duality \cite{Klebanov:2000hb}. We can understand the cascade by following $b_0$: everytime we lower the radial coordinate (and thus we lower the energy scale) such that $b_0 \to b_0-1$, we have descent one step of the cascade. Then we can look at the shift of Maxwell charges in this process.

In our solution the functions are such that in the IR we can neglect $\tilde p/M$ with respect to $f$. Then in one cascade step we experience:
\begin{equation}
f^{(i)} \to f^{(i-1)} = f^{(i)}-\frac{2\pi}{M}
\qquad\quad \Rightarrow \qquad\quad \begin{aligned}
M_{eff}^{(i)} &\to M_{eff}^{(i-1)} \simeq M_{eff}^{(i)} - N_f \\
N_{eff}^{(i)} &\to N_{eff}^{(i-i)} \simeq N_{eff}^{(i)} - M_{eff}^{(i)} + \frac{N_f}{2} \;.
\end{aligned}
\end{equation}
Here $i$ is an integer that counts the number of Seiberg dualities from the bottom up. Without taking the IR limit, both $M_{eff}(\rho)$, $N_{eff}(\rho)$ and $b_0(\rho)$ are positive monotonically increasing functions, of which I give the IR and UV expansions:
\begin{equation}
IR: \; \begin{aligned}
M_{eff} &= \frac{MN_f}{2\pi} \, \frac{1}{\rho^2} + \orders{\frac{1}{\rho^3}} \\
N_{eff}-N_0 &= \frac{M^2N_f}{8\pi^2} \, \frac{1}{\rho^4} + \orders{\frac{1}{\rho^5}}
\end{aligned}
\qquad\quad UV: \; \begin{aligned}
M_{eff} &= \frac{MN_f}{2\pi} \, \frac{\alpha}{2(-\rho)} + \calO(\rho) \\
N_{eff}-N_0 &= \frac{M^2N_f}{8\pi^2} \, \frac{\alpha^2}{2\rho^2} + \orders{\frac{1}{\rho}}
\end{aligned} \end{equation}

These formulae are not satisfactory at all. First of all they are only approximate; moreover in the UV the functions $f$ and $\tilde p$ are of the same order giving a very different result, and in the middle there in no clear pattern. The reason is that we are looking at the wrong objects. As fully explained in \cite{Benini:2007gx}, Maxwell charges are gauge invariant and conserved, but are not quantized nor localized: they gain contributions from the whole bulk and from the charges induced on the D7-branes. Thus they are not suitable for identifying gauge ranks. The correct objects to look at are Page charges: they are quantized and localized on the D3 and D5-branes that source them (they are not even sourced by the induced charges on the D7's). On the other hand they are not invariant under large gauge transformations. These ones, which are quantized themselves, precisely correspond to Seiberg dualities and we expect Page charges to change accordingly.

\subsection{Chiral Zero Modes}

Before going on with the computation of Page charges, I want to give a physical explanation of the origin of the chiral gauge singlet fields $\Phi_j$ transforming in the $(\overline{N_f},N_f)$ flavor representation. For this, we need to do a little digression.

Consider the following brane configuration: put two stacks of $N_f$ spacetime filling intersecting D7-branes on $M_{3,1}\times T^6$ (I am interested in the local physics, so I neglect tadpole cancellation issues). Each stack wraps a $T^4$ in $T^6$ and they intersect along a $T^2$ (times Minkowski spacetime). The theory at the intersection is an \Nugual{1} 6d chiral gauge theory with 8 supercharges, gauge group $U(N_f)\times U(N_f)$ and bifundamental chiral matter. Of course, when compactified to 4d, the theory is \Nugual{2} non-chiral. Moreover in the decompactification limit and from a 4d point of view, the whole theory gets frozen (being higher dimensional).

The situation is different if we put some (supersymmetric) gauge flux on the D7-branes. The number of supercharges is reduced to 4, signaling that a 4d dynamics is taking place. This is in fact the case, as the system is T-dual to D6-branes in Type IIA intersecting at angles. Due to the non-trivial flux $F_2$ in IIB, the D6-branes intersect at non-right angles on all of the six directions; the intersection is four dimensional and 4d chiral modes arise there, transforming in the bifundamental representation. The honest computation in IIB was performed in \cite{Cremades:2004wa} (actually in the context of magnetized D9-branes). The net effect of the flux, which is pulled-back to the intersection, is to twist the Dirac operator so that there are a number of zero modes. This number is given by the difference between the fluxes on the stacks:
\begin{equation}
N_\Phi = \frac{1}{2\pi} \int_{T^2} (F_2^{(A)} - F_2^{(B)}) \;.
\end{equation}
In \cite{Cremades:2004wa} it was also shown that the zero modes are localized at a point in the 6d intersection, developing a 4d identity. This obviously corresponds to the intersection being four dimensional in IIA. Moreover, in the decompactification limit the gauge theory decouples but the zero modes preserve their 4d essence. As the fluxes on the D7's are quantized so is the number of zero modes, which corresponds to the number of intersections in IIA.

In our setup we have a very similar situation. We have two stacks of D7-branes%
\footnote{After the smearing all the branes in a stack are separated, that means that the gauge theory on them is in the Coulomb phase and $U(N_f)$ is broken to $U(1)^{N_f}$. This does not change the conclusion.}
which intersect along an holomorphic submanifold of complex dimension 1 and with topology of $\bbC^*$. On the branes there are opposite gauge fluxes, which I expect giving rise to chiral zero modes with 4d dynamics and transforming in the $(\overline{N_f},N_f)$ representation of the flavor group. Unfortunately the intersection is non-compact thus an equally clean derivation is not possible. Nevertheless, in our supersymmetric setup (where charges are equal to masses so that they always sum and never cancel together) we can interpret $F_2$ as providing a density of zero modes. This means that integrating $F_2$ on a region we get the number of zero modes originating from there.%
\footnote{The actual position of the zero modes is encoded in the Wilson lines of the gauge connection. So in principle the zero modes could be everywhere, not necessarily in the region we are considering.}

We take the gauge field strength in our solution \eqref{fieldstrength F2} and pull-back $F_2^{(\Sigma_2)} - F_2^{(\Sigma_1)}$ on the intersection $\Pi=\Sigma_1\cap\Sigma_2$. We get:
\begin{equation}
2\pi F_{int} \equiv ( 2\pi F_2^{(\Sigma_2)} - 2\pi F_2^{(\Sigma_1)} ) \Bigr|_\Pi = \frac{2}{3} e^{2u} p(\rho)\, d\rho\wedge d\psi \;.
\end{equation}
Notice that in the far IR the gauge field strength on the branes goes to zero, confirming that the IR field theory does not have extra gauge singlet fields. Then we produce a function that counts the number of zero modes from the far IR $\rho = - \infty$ to some energy scale $\rho$ by integrating the gauge field strength $F_{int}$ on the intersection $\Pi$ up to the radius $\rho$:
\begin{equation}
N_\Phi(\rho) = \frac{1}{4\pi^2} \int_{\Pi[-\infty,\rho]} 2\pi F_{int} = \frac{M}{2\pi} f(\rho) - \frac{1}{3\pi} \tilde p(\rho)
\end{equation}

Now we can perform an IR analysis in the region $|\rho| \gg 1$. Neglecting the function $\tilde p/M$ with respect to $f$, in the shift $f(\rho) \to f(\rho-\Delta\rho) = f(\rho) -2\pi/M$ which corresponds to one Seiberg duality towards the IR we have a shift
\begin{equation}
N_\Phi(\rho) \to N_\Phi(\rho-\Delta\rho) \simeq N_\Phi(\rho)-1 \;.
\end{equation}
This result confirms that, at least in the IR, in each Seiberg duality we lose one chiral zero mode $\Phi$ in the bifundamental flavor representation.

It would be nice to give an interpretation to the scaling of $N_\Phi$ in the UV. Moreover it would be interesting to give a more rigorous counting of the zero modes contained in the throat up to some radius (energy scale) $r_0$; a possible solution could be appealing to the index theorem with boundary.%
\footnote{I thank B. Acharya and G. Shiu for this suggestion.}
I leave these issues for future investigations.

\subsection{Page Charges}

Page dual currents \cite{Page:1984qv} can be obtained by writing the Bianchi identities with sources as total differentials. The only terms that cannot be written in this way are the source delta functions corresponding to the D3 and fractional D3-branes at the tip of the conifold that produce our background, and that are replaced by their fluxes in the geometric transition. In particular the Page charges obtained by integration do not get contributions from the bulk nor from the induced charges on the D7-branes, are independent of the radial coordinate where we measure them and are quantized, making them very suitable to measure gauge ranks.

In general $b_0$ takes in the far IR some limiting value $b_2^{(0)}$, that we conventionally choose in the range $b_2^{(0)} \in [0,1]$. This range is special because it returns us positive square gauge couplings when exploiting usual formulae \cite{Herzog:2002ih}. Then, moving towards the UV, $b_0$ starts growing, ending up out of that range at a generic energy scale. We could say that the field theory is still the one of the IR, but such a description is not useful because the gauge couplings have grown diverging and then becoming imaginary. Thus we had better shift $b_0$ by $-n$ units bringing it back to the range $[0,1]$; this process is a large gauge transformation or a Seiberg duality. We end up with a new equivalent field theory description, with different gauge ranks but real positive gauge couplings. In this way making large gauge transformations at a fixed energy scale (which changes the Page charges) is a way of understanding the cascade.

Our Page dual currents are
\begin{align}
\ast j_{D5}^{Page} &= F_3 - B_2\wedge F_1 + 2\pi A \wedge \Omega_2 \\
- \ast j_{D3}^{Page} &= F_5 - B_2\wedge F_3 + \frac{1}{2}B_2\wedge B_2\wedge F_1 + \frac{1}{2} 2\pi A\wedge 2\pi dA\wedge \Omega_2 \;.
\end{align}
One can check they are in fact closed forms. Page charges are obtained by integrating their differentials:
\begin{equation} \begin{aligned}
Q_{D5}^{Page} &= \frac{1}{4\pi^2\alpha'} \int_{V_4} d\ast j_{D5}^{Page} \\
Q_{D3}^{Page} &= \frac{1}{(4\pi^2\alpha')^2} \int_{V_6} d\ast j_{D3}^{Page} \;,
\end{aligned} \end{equation}
where $V_4$ and $V_6$ are bounded by $S^3$ and $T^{1,1}$. Using Stoke's theorem we eventually get:
\begin{equation} \begin{aligned} \label{Page charges}
Q_{D5}^{Page} &= \frac{1}{4\pi^2\alpha'} \int_{S_3} \Bigl( F_3 - B_2\wedge F_1 + 2\pi A \wedge \Omega_2 \Bigr) \\
Q_{D3}^{Page} &= -\frac{1}{(4\pi^2\alpha')^2} \int_{T^{1,1}} \Bigl( F_5 - B_2\wedge F_3 + \frac{1}{2}B_2\wedge B_2\wedge F_1 + \frac{1}{2} 2\pi A\wedge 2\pi dA\wedge \Omega_2 \Bigr) \;.
\end{aligned} \end{equation}

We compute the Page charges of our solution. Some care is needed in the evaluation of the smeared forms (see the discussion at page \pageref{discussion smear}). One gets
\begin{equation} \begin{aligned} \label{Page charges far IR}
Q_{D5}^{Page} &= -N_f \, b_2^{(0)} \\
Q_{D3}^{Page} &= N_0 + \frac{N_f}{2} \, (b_2^{(0)})^2 \;.
\end{aligned} \end{equation}
After identifying a dictionary between supergravity and field theory, we will match these charges with the IR of the theory.

Then I am interested in how these quantities change under a large gauge transformation of $B_2$. We perform $B_2 \to B_2 + \Delta B_2$ with
\begin{equation}
\Delta B_2 = -n\pi \, \omega_2 \qquad\qquad n\in \bbZ \;.
\end{equation}
It is a gauge transformation because $\Delta H_3 = 0$ and $\frac{1}{4\pi^2} \int_{S^2} \Delta B_2 = -n$ ($n$ identifying the number of Seiberg dualities) and is large because $\Delta B_2$ is not an exact form. A shift of $B_2$ must be accompanied by a shift of the gauge connection $A$ on the branes, since $\calF$ is the gauge invariant quantity. Thus $2\pi d\Delta A = -\Delta \hat B_2$. We find
\begin{equation}
2\pi d\Delta A \Bigr|_{\Sigma_2} = n \frac{\pi}{2} \sin\theta_1\, d\theta_1\wedge d\varphi_1 \qquad\qquad
2\pi \Delta A \Bigr|_{\Sigma_2} = -n \frac{\pi}{2} \hat g^5 \;.
\end{equation}
The variations on $\Sigma_1$ are the same but with opposite sign.

The variation of the D5 Page charge is readily obtained: $\Delta Q_{D5}^{Page} = n N_f$. In the computation of the D3-charge I imagine having already shifted $B_2$ by $m$ units, so that we use:
\begin{equation}
B_2 = \Bigl( \frac{M}{2} f + (b_2^{(0)} - m)\, \pi \Bigr)\, \omega_2 \qquad\qquad 2\pi A \Bigr|_{\Sigma_2} = \Bigl( \frac{M}{4} \, f  - \frac{1}{6} \, \tilde p + \frac{\pi}{2} \, (b_2^{(0)} - m) \Bigr)\, \hat g^5 \;.
\end{equation}
After some algebra we get
\begin{equation}
\Delta Q_{D3}^{Page} = n\, (m-b_2^{(0)}) \, N_f + n^2 \frac{N_f}{2} \;.
\end{equation}
Notice that the first piece is the D5-charge before the shift.

We can summarize here the result:
\begin{equation} \label{Page charges shift}
\left\{ \begin{aligned}
\Delta Q_{D5}^{Page} &= n\, N_f \\
\Delta Q_{D3}^{Page} &= n\, Q_{D5}^{Page} + n^2\, \frac{N_f}{2} \;.
\end{aligned} \right.
\end{equation}
The formula is consistent with subsequent shifts and with \eqref{Page charges far IR}. The case $n=-1$ corresponds to one Seiberg duality towards the IR. Notice that it gives the same approximate IR result derived with Maxwell charges. Anyway Page charges give us an exact and much cleaner result.

Unfortunately I was not able to find a kind of Page charge to measure the number of chiral zero modes. Let me just notice that the quantity $N_\Phi$ does not change under the large gauge transformations we considered. I leave this problem for the future.

\section{Brane Engineering} \label{sec:brane engineering}

In this section I engineer the effective field theory at some energy scale with probe branes on the singular conifold, and compute the charges generated by such a configuration. In this way one can construct a dictionary between the supergravity (Page) charges and the field theory ranks. Initially the goal is to construct a generic theory with gauge group $SU(g_1)\times SU(g_2)$ ($g_1 \geq g_2$), flavor group $U(N_f)\times U(N_f)$ and $k$ gauge singlet fields in the $(\overline{N_f},N_f)$ flavor representation. As we will see this is not easy, and we will restrict to the class of non-anomalous theories. Nonetheless this is enough to understand the cascade.

The gauge theory is realized as the near horizon theory on a stack of fractional D3-branes, which can be thought as D5-branes wrapped on the 2-cycle of $T^{1,1}$ and possibly with gauge flux on them. The computation is as in \cite{Grana:2001xn,Benini:2007gx}. The Wess-Zumino action for a D5-brane with flux is
\begin{equation}
S_{D5} = \mu_5 \int_{M_{3,1}\times S^2} \Bigl[ \hat C_6 - (\hat B_2 + 2\pi F_2) \wedge \hat C_4 \Bigr] \;.
\end{equation}
We consider a flat background value for $B_2$ proportional to $\omega_2$, and $F_2$ can be expanded on the pull-back of $\omega_2$ on the brane:
\begin{equation} \label{expansion fluxes}
B_2 = \pi b_0 \, \omega_2 \qquad\qquad\qquad 2\pi F_2 = \pi \phi_0 \, \hat \omega_2 \;,
\end{equation}
so that $\frac{1}{4\pi^2} \int_{S^2} B_2 = b_0$ and $\frac{1}{4\pi^2} \int_{S^2} 2\pi F_2 = \phi_0$. The gauge bundle is quantized according to $\phi_0 \in \bbZ$. We read that the D5-charge is $1$ and the D3-charge is $-(b_0 + \phi_0)$. For an anti-D5-brane the charges are the opposite: D5-charge $-1$ and D3-charge $(b_0 + \phi_0)$. The gauge theory of interest is realized with $g_1$ D5-branes and $g_2$ anti-D5's with one unit of flux. The charges are summarized in Table \ref{tab:dictionary}.

Then we consider the case of a D7-brane without flux, with one branch along $\Sigma_1 =\{\theta_1,\varphi_1 = \text{const}\}$ and one along $\Sigma_2 =\{\theta_2,\varphi_2 = \text{const}\}$. The Wess-Zumino action is:
\begin{equation} \label{probe D7 action}
S_{D7} = \mu_7 \int_{M_{3,1}\times \Sigma} \Bigl[ C_8 - (\hat B_2 + 2\pi F_2) \wedge C_6 + \frac{1}{2} (\hat B_2 + 2\pi F_2)^2 \wedge C_4 - \frac{\pi^4}{3} p_1(\calR)\wedge C_4 \Bigr] \;.
\end{equation}
The topology of the branches in the singular conifold is $\bbC^2 \setminus \{0\}$ because they pass through the singularity and a resolution is needed in order to understand the physics. In the resolution of the conifold one branch participate to the blowing up of the 2-cycle, giving rise to $\widehat{\bbC^2}$ ($\bbC^2$ blown up at a point), while the other one is not modified and only touches the exceptional cycle at a point (see Appendix \ref{sec:exceptional}). Which one of the branches is blown up is reversed by a flop transition, anyway the physics does not depend on this choice.

The case without flux is the one considered in \cite{Benini:2006hh}, and the case we expect to be realized in the far IR of our solution. Even if we are not putting flux on the brane, we cannot just take $F_2=0$ because the pull-back of $B_2$ does not go to zero at infinity and one would get an infinite induced charge. Thus we set an $F_2$ that kills the tail of $\hat B_2$ at infinity but has no flux on $S^2$ (in the $\widehat{\bbC^2}$ case). The resulting $\calF$ is zero on the $\bbC^2$ branch and is the Poincar\'e dual to $S^2$ on the $\widehat{\bbC^2}$ branch. The details of this computation are in Appendix \ref{sec:exceptional}.

I call $\sigma_2$ the Poicar\'e dual to $S^2$ on the $\widehat{\bbC^2}$ branch; it satisfies
\begin{equation}
\int_{S^2} \alpha_2 = \int \alpha_2 \wedge \sigma_2 \qquad\qquad \int \sigma_2\wedge \sigma_2 = - \,\#(S^2,S^2) = 1
\end{equation}
for every (normalizable) closed 2-form $\alpha_2$. $-1$ is the self-intersection of $S^2$ on $\widehat{\bbC^2}$.%
\footnote{The minus sign in front of the intersection form comes from my conventions on the volume forms on $\Sigma$ and $S^2$, both inverted (see Appendix \ref{sec:conventions} and \ref{sec:exceptional}).}
Thus the gauge fluxes on the two branches are:
\begin{equation}
\calF \Bigr|_{\bbC^2} = 0 \qquad\qquad\qquad \calF \Bigr|_{\widehat{\bbC^2}} = 4\pi^2 b_0 \, \sigma_2 \;.
\end{equation}
Then the two reduced actions are:
\begin{align}
S_{D7}(\bbC^2) &= \mu_7 \int_{M_{3,1}\times \bbC^2} C_8 + (\text{curv}) \, \mu_3 \int_{M_{3,1}} C_4 \label{reduced D7 action C2}\\
S_{D7}(\widehat{\bbC^2}) &= \mu_7 \int_{M_{3,1}\times \widehat{\bbC^2}} C_8 - b_0\, \mu_5 \int_{M_{3,1}\times S^2} C_6 + \Bigl[ \frac{b_0^2}{2} + (\text{curv}) \Bigr] \mu_3 \int_{M_{3,1}} C_4 \;. \label{reduced D7 action C2 hat}
\end{align}
The curvature couplings are computed in Appendix \ref{sec:exceptional} for completeness, even if they do not play an important r\^ole.
The induced charges can be immediately read from these expressions, and are summarized in Table \ref{tab:dictionary}.
This result is in perfect agreement with the Page charges \eqref{Page charges far IR} in the far IR, confirming that our theory flows to the flavored KW theory.

At this point we can readily obtain the charges sourced by a D7-brane with flux as well. Obviously we can only put some $F_2$ flux on the $\widehat{\bbC^2}$ branch, since the other one does not have any 2-cycle. To add $\phi_0$ units of $F_2$ flux on $S^2$ we substitute $b_0$ with $(b_0+\phi_0)$ in the expression of $\calF$. Again the result is in Table \ref{tab:dictionary}.

\begin{table}[t]
\begin{center}
\begin{tabular}{|l|c|c|c|}
\hline \tabs
          & frac D3$_{(1)}$ & frac D3$_{(2)}$ & D7$^{\Sigma_1}$ + D7$^{\Sigma_2}$ \\
\hline \hline \tabs
D3-charge & $-b_0$   & $1+b_0$ & $\frac{1}{2} (b_0 + \phi_0)^2 + (\text{curv})$ \\
\tabs
D5-charge & $1$      & $-1$    & $-(b_0+\phi_0)$ \\
\tabs
D7-charge & 0        & 0       & 1+1 \\
\hline
Number of objects & $g_1$ & $g_2$ & $N_f$ \\
\hline
\end{tabular}
\caption{Effective charges for fractional D3-branes and D7-branes with flux. \label{tab:dictionary}}
\end{center}
\end{table}

\

One could think that the number $\phi_0$ of units of flux on the D7-branes corresponds to the number $N_\Phi$ of zero modes arising at the intersection, thus to the number of gauge singlets in field theory. Actually this is not exact. The reason is that for generic values of the gauge ranks and of the number of gauge singlets, the chiral flavor symmetry is anomalous. From the gravity point of view, the action of the gauge theory living on the D7's is not gauge-invariant; the variation is a boundary term, and since the branes are non-compact this is not an inconsistency and only represents an anomaly for a global symmetry in field theory.

There are two kinds of possible sources of anomaly. The first one arises as a would-be tadpole on the D7-branes: since $d\calF = \hat H_3$, if the cohomology class of $\hat H_3$ on the 4-cycle is non-vanishing there is a tadpole.
% \cite{Freed:1999vc}.
In our case $\int_{\calC_3} H_3 =0$ for every compact 3-submanifold on the D7 worldvolume so that there are no tadpoles. The second one is precisely the anomaly for the chiral flavor symmetry.
It could be computed by performing a gauge variation $\delta A = d\lambda$ of the Wess-Zumino action for a D7-brane, along the lines of \cite{Green:1996dd, Witten:1998zw} and more recently \cite{Casero:2007ae}. An anomaly is seen as a non-vanishing variation of the boundary term, so that the absence of f-f-f anomalies translates into $\delta_\lambda S_{WZ} = 0$.

Thus suppose starting with a configuration of $N_0$ D3-branes and $N_f$ D7-branes without flux, which is the non-anomalous flavored KW theory. We can put one unit of flux ($\phi_0 = -1$) on each $\widehat{\bbC^2}$ branch of D7. This gives us a new non-anomalous configuration.
%, because the addition to $\calF$ of a closed $2\pi F_2$ does not change $[\hat H_3]$.
From Table \ref{tab:dictionary}, the modification of the charges is that of the addition of $N_f$ D5-branes wrapped on $S^2$ and a D3-charge of $\frac{N_f}{2}$. On the other hand, we know that a non-trivial cohomology class $2\pi F_2$ for the D7 gauge bundle represents D5-branes dissolved (or even localized) into the D7's. In particular a flux on $S^2$ represents D5-branes that wrap the 2-cycle.%
\footnote{Notice that since one of the D7 branches wraps the shrunk 2-cycle, the D5-branes wrapped on it must necessarily lie inside the D7.}
The new non-anomalous theory is thus engineered by $N_0$ D3-branes, $N_f$ D5-branes and $N_f$ D7-branes, and being non-anomalous there must be one gauge singlet field. What we have found is precisely our field theory at the first (from the bottom) step of the cascade.

This is a general pattern. Each unit of flux on the D7's corresponds to the addition of $N_f$ fractional D3-branes (thus increasing the difference of the gauge ranks by $N_f$), one gauge singlet field to preserve the anomaly and a number of $D3$ branes. We will match this pattern with the field theory cascade in the next section. If we want to isolate the charge contribution of one gauge singlet field, it is just a D3-charge of $\frac{N_f}{2}$. In Table \ref{tab:dictionary 2} we report this different counting of charges.

\begin{table}[t]
\begin{center}
\begin{tabular}{|l|c|c|c|c|}
\hline \tabs
      & frac D3$_{(1)}$ & frac D3$_{(2)}$ & D7$^{\Sigma_1}$ + D7$^{\Sigma_2}$ & $N_\Phi$  \\
\hline \hline \tabs
D3-charge & $-b_0$   & $1+b_0$ & $\frac{1}{2} b_0^2 + (\text{curv})$ & $\frac{N_f}{2}$ \\ 
\tabs
D5-charge & $1$      & $-1$    & $-b_0$                              & 0 \\
\tabs
D7-charge & 0        & 0       & 1+1                                 & 0 \\
\hline \tabs
Number of objects & $g_1$ & $g_2$ & $N_f$ & $k$ \\
\hline
\end{tabular}
\caption{Effective charges for fractional D3-branes, D7-branes without flux and $N_\Phi$ gauge singlets. \label{tab:dictionary 2}}
\end{center}
\end{table}

% \
%
% I can bring another weaker argument for the charge assignment of $\frac{N_f}{2}$ to $N_\Phi$. This field transforms in the $(\overline{N_f},N_f)$ flavor representation so it is made of $N_f^2$ singlet fields. A way of introducing 4d singlet fields in our theory without increasing any gauge rank is to add D3-branes on the D7's. We put the D3-branes out of the singularity, so the 3-3 strings charged under the gauge group are massive. Moreover when the D3-branes are inside the D7's, the gauge theory on them is in the Higgs branch and the D3's do not provide new gauge groups. They provide instead

\subsection{The Cascade} \label{sec:the cascade}

I conclude with the matching of the cascade between field theory and supergravity. We consider at step $(i)$ a theory with gauge group $SU(g_1)\times SU(g_2)$ (with $g_1 > g_2$), flavor group $U(N_f)\times U(N_f)$ and $k$ gauge singlet fields in the $(\overline{N_f},N_f)$ flavor representation. It is realized with $g_1$ fractional D3-branes of type one and $g_2$ of type two. The Page charges sourced by this configuration are (Table \ref{tab:dictionary 2}):
\begin{equation} \begin{aligned}
M^{(i)} &= g_1 - g_2 - b_0 N_f \\
N^{(i)} &= -b_0\, g_1 + (1 + b_0)g_2 + \frac{b_0^2}{2} + \frac{N_f}{2} k + (\text{curv}) \;.
\end{aligned} \end{equation}
After one Seiberg duality towards the IR we have at step $(i-1)$ a theory with gauge group $SU(g_2)\times SU(2g_2 +N_f-g_1)$, the same flavor group and $k-1$ gauge singlet fields. The new Page charges are:
\begin{equation} \begin{aligned}
M^{(i-1)} &= g_2 - (2g_2 + N_f - g_1) - b_0 N_f \\
N^{(i-1)} &= -b_0\, g_2 + (1 + b_0)(2g_2 + N_f - g_1) + \frac{b_0^2}{2} + \frac{N_f}{2} (k-1) + (\text{curv}) \;.
\end{aligned} \end{equation}
Thus we verify that
\begin{equation}
\left\{ \begin{aligned}
M^{(i-1)} &= M^{(i)} - N_f \\
N^{(i-1)} &= N^{(i)} - M^{(i)} + \frac{N_f}{2} \;,
\end{aligned} \right.
\end{equation}
in perfect agreement with the supergravity computation \eqref{Page charges shift}.

\

The careful reader could wonder what is the r\^ole of the constant $M$ in the supergravity solution. In the field theory there is no rank controlled by it: in the IR the gauge ranks are equal and controlled by $N_0$; then, going towards the UV, at some energy scale they start growing and the cascade is controlled by $N_f$. The parameter $M$ does not enter, and in fact it is not even quantized.

It turns out that $M$ fixes the energy scale of the last (lower) Seiberg duality. This last step (after which the theory does not cascade any more) takes place at a radius $r_0$ such that $b_0(r_0) = \frac{1}{4\pi^2} \int_{S^2}B_2$ is $1$. Then, negletting for clarity $b_2^{(0)}$, one finds $f(r_0) = 2\pi/M$. The biggest is $M$, the smaller is the energy scale of the last Seiberg duality compared with the duality wall scale, and the larger is the number of dualities contained in the weakly coupled supergravity description.

\section{Conclusions}

In this paper I presented a field theory obtained as a chiral flavoring of the Klebanov-Tseytlin theory. The RG flow is understood as a cascade of Seiberg dualities in which flavors actively participate, and new gauge singlet fields have to be taken into account. Then I proposed a gravity dual, constructed by putting backreacting flavor D7-branes with flux in a background. The existence of a gravity dual gives more sturdy ground to the cascade, and allows us to predict the full non-perturbative RG flow.

The UV theory presents a duality wall as well as a Landau pole, as it happens in \cite{Benini:2007gx}. The fact that $b_0(\rho)$ diverges as approaching the Landau pole tells us that an infinite number of Seiberg dualities would be necessary to reach a finite energy scale, and the number of degrees of freedom diverges as well. Of course this has to be taken with a grain of salt as the string coupling (and the gauge coupling) diverges as well. On the other side, along the cascade the difference between the gauge ranks reduces going towards the IR. At some point they get equal and there is no cascade any more. The string coupling always decreases, which initially translates into both gauge groups having positive $\beta$-function and the gauge couplings flowing towards zero. As explained in \cite{Benini:2006hh}, at some point $g_s N_f$ becomes small, the flavor branes do not backreact any more and the gravity solution asymptotes the KW one but with smaller and smaller string coupling. On the field theory side the gauge coupling stops at some minimal value $g_*$ (the extremum of the line of conformal points, where the quartic superpotential vanishes) and what still flows to zero is the flavor superpotential coupling. Eventually the theory reaches a fixed superconformal point with flavors, vanishing superpotential and gauge coupling $g_*$. This is badly described by supergravity.

There are a number of comments in order. Computing Maxwell D-brane charges and Page charges, we discovered the first instance where they are not on the same footing: the cascading of gauge ranks is perfectly described by Page charges but \emph{not} by Maxwell charges. It would be interesting to understand which is instead the physical information encoded in Maxwell charges.

The flavoring of the KT cascade is interesting for another reason. When trying to generalize it to fractional branes at more generic conical singularities (see \cite{Herzog:2004tr} for an example), an IR problem arises: if there are no complex deformations the singularity cannot be resolved, the field theory presents a runaway behavior and/or it breaks supersymmetry \cite{Berenstein:2005xa}. The addition of flavor branes can cure this problem, as fractional branes can disappear in the IR and the field theory still flows to a superconformal point. When trying to flavor these theories with D7-branes one discovers that generically it is not possible to do it in the non-chiral way of \cite{Benini:2007gx}. The flavors generically couple to operators with non-zero baryonic numbers; on the gravity side, generically the pull-back of $H_3$ on the 4-cycles is different from zero. Thus the most general situation is the one exemplified in this paper.

For instance the authors of \cite{Franco:2006es} used the last step of a cascade obtained by flavoring the cone over $dP_1$ (equivalently $Y^{2,1}$) to study realizations of the ISS mechanism \cite{Intriligator:2006dd} in string theory. The flavoring they consider is of chiral type, with a cascade quite similar to the one presented here. It would be interesting to explicitly realize the gravity duals to those models.

Lastly, the appearance of 4d chiral zero modes along the intersection of branes with flux could have a relevance for the construction of phenomenological models. In \cite{Acharya:2006mx} it was considered a mechanism for localizing fermions in the bulk of a Randall-Sundrum throat. Here we explicitly see another possibility.

\section*{Acknowledgments}

I would like to warmly thank Bobby Acharya, Matteo Bertolini, Giulio Bonelli, Felipe Canoura, Stefano Cremonesi, Jose Edelstein, Sameer Murthy, Carlos Nu\~nez, Alfonso Ramallo, Gary Shiu, Angel Uranga, Roberto Valandro and David Vegh for discussions, suggestions and criticism. A particular thanks goes to Carlos Nu\~nez and Albero Zaffaroni for reading the manuscript and giving many useful comments. Lastly I would like to thank the University of Santiago de Compostela for hospitality during the early stage of this work.

\appendix

\section{Conventions and Calabi-Yau Geometry} \label{sec:conventions}

My conventions on the Hodge dual in six and ten dimensions is that $F_p \wedge \ast F_p = |F_p|^2 \, \text{vol}_n$, where $|F_p|^2 = \frac{1}{p!} (F_p)_{\mu_1\dots\mu_n} (F_p)_{\nu_1\dots\nu_n} g^{\mu_1\nu_1} \dots g^{\mu_n\nu_n}$. Then the Euclidean Hodge dual in 6d and the Lorentzian one in 10d (mostly plus signature) act on a vielbein basis respectively as:
\begin{align}
\ast_6 (e^{a_1}\wedge\dots\wedge e^{a_p}) &= \frac{\epsilon^{a_1\dots a_p b_1\dots b_{n-p}}}{(n-p)!} \,  e^{b1}\wedge\dots\wedge e^{b_{n-p}} \\
\ast_{10} (e^{a_1}\wedge\dots\wedge e^{a_p}) &= \frac{\epsilon^{a_1\dots a_p b_1\dots b_{n-p}}}{(n-p)!} \, (-1)^{\delta_{a_i,0}}\,  e^{b1}\wedge\dots\wedge e^{b_{n-p}} \;,
\end{align}
where $\delta_{a_i,0}$ is 1 only when one of the $a_i$ is the time component. The 6d K\"ahler form and holomorphic form \eqref{Kahler form} and \eqref{holomorphic form} satisfy the following properties: are closed, co-closed, $J\wedge J=2\ast_6 J$, $\ast_6\Omega=-i\Omega$ and $J^3 = \frac{3i}{4} \Omega\wedge \overline{\Omega}=6 \text{vol}_6$. A convenient vielbein basis, compatible with metric, complex structure and orientation, is:
\begin{equation}
dr,\quad \frac{1}{3}r\, g^5,\quad \frac{r}{\sqrt{6}}d\theta_1,\quad -\frac{r}{\sqrt{6}}\sin\theta_1\, d\varphi_1,\quad \frac{r}{\sqrt{6}}d\theta_2,\quad -\frac{r}{\sqrt{6}}\sin\theta_2\, d\varphi_2 \;.
\end{equation}
Notice that the induced volume forms on the 4-cycles $\Sigma_j$ and the 2-cycle $S^2$ get a minus sign in front.

The holomorphic (3,0)-form can be derived in the following way. First the complex structure of the conifold is inherited from $\bbC^4$ by defining it as the locus $z_1z_2 - z_3z_4 = 0$. The coordinates $z_i$ are related to my coordinates through
\begin{equation} \begin{aligned}
z_1 &= r^{3/2} e^{i/2(\psi-\varphi_1-\varphi_2)} \sin(\theta_1/2) \sin(\theta_2/2) \qquad &
z_3 &= r^{3/2} e^{i/2(\psi+\varphi_1-\varphi_2)} \cos(\theta_1/2) \sin(\theta_2/2) \\
z_2 &= r^{3/2} e^{i/2(\psi+\varphi_1+\varphi_2)} \cos(\theta_1/2) \cos(\theta_2/2) &
z_4 &= r^{3/2} e^{i/2(\psi-\varphi_1+\varphi_2)} \sin(\theta_1/2) \cos(\theta_2/2) \;.
\end{aligned} \end{equation}
We can perform the linear change of coordinates
\begin{equation}
z_1 = w_1 + i\,w_2 \qquad z_2 = w_1 - i\,w_2 \qquad z_3 = w_3 + i\,w_4 \qquad z_4 = -w_3 + i\,w_4 \;,
\end{equation}
so that the conifold equation becomes $w_1^2 + w_2^2 + w_3^2 + w_4^2 =0$. Then the holomorphic form is given by
\begin{equation}
\Omega = \frac{dw_2\wedge dw_3\wedge dw_4}{w_1} \;.
\end{equation}

\section{Equations of Motion} \label{sec:EOM}

The equations of motion for the form-fields can easily be derived from the Einstein frame action in the following fashion:
\begin{equation} \begin{aligned}
S_{IIB} &= \frac{1}{2\kappa_{10}^2} \biggl\{ \int d^{10}x \, \sqrt{-g} \, R \; -\frac{1}{2} \int \Bigl[ \partial\phi\wedge\ast\partial\phi + e^{2\phi} F_1\wedge\ast F_1 + \frac{1}{2} F_5\wedge\ast F_5 \\
&\qquad\qquad\qquad\qquad\qquad\qquad  + e^{-\phi} H_3\wedge\ast H_3 + e^\phi F_3\wedge\ast F_3 + C_4\wedge H_3 \wedge F_3 \Bigr] \\
&\quad - \int d^8\xi \, e^\phi \, \sqrt{-\det(\hat g+ e^{-\phi/2}\, \calF)} \\
&\quad +\int \Bigl[ \hat C_8 - \hat C_6\wedge \calF + \frac{1}{2} \hat C_4\wedge \calF \wedge \calF - \frac{1}{6} \hat C_2 \wedge \calF^3 + \frac{1}{24} C_0 \, \calF^4 \Bigr] \biggr\} \;.
\end{aligned} \end{equation}
I substituted $2\kappa_{10}^2\mu_7=1$.

Unfortunately, not all the source terms are correctly derived from it. For instance, the source term in $dF_5$ comes without the $\frac{1}{2}$ factor in front. The same problem arises if we consider a D3-brane: the action $S_{WZ}=\mu_3 \int \hat C_4$ gives us twice the correct source term in the BI of $F_5$. The origin of this is that a D3-brane is both an electric and magnetic source, and a Lagrangian formulation is not suitable for it. A good guiding principle is requiring the full system of EOM's to be consistent with $d^2=0$.

Problematic is also the EOM for $d(e^{-\phi}\ast H_3)$. The bulk computation involves in general all gauge potentials, which are not defined in the presence of the sources we want to take into account. Thus our strategy will be that of introducing sources one-by-one. At the level of EOM's, all the sources can be introduced at the same time.

The variation of the type IIB bulk action with respect to $B_2$ is:
\be \label{B2 generating equation}
2\kappa_{10}^2 \delfrac{S_\text{IIB}}{B_2} = \frac{1}{2}\, d\Big( -C_2 \wedge *F_5 + 2\, e^{-\phi} * H_3 - 2\, C_0\, e^\phi *F_3 + C_4 \wedge F_3 + C_0\, C_4 \wedge H_3 \Big) \;.
\ee
Without sources, the EOM is easily worked out substituting the BI's, and we get (\ref{modified BI EOM}). Anyway, in the presence of sources there are obstructions. If $dF_p - H_3\wedge F_{p-3} \neq 0$ then $C_{p-1}$ is not defined. In these situations the equation is meaningless.
Then we perform a partial integration in (\ref{B2 generating equation}), exploiting that $d^2=0$:
\be
2\kappa_{10}^2 \delfrac{S_\text{IIB}}{B_2} = \frac{1}{2}\, d\Big( 2\, e^{-\phi} *H_3 - H_3 \wedge C_2 \wedge C_2 - 2\, C_2 \wedge F_5 + 2\, C_0\, F_7 \Big)
\ee
so that the equation is:
\begin{multline} \label{H3 equation 1}
d\big( e^{-\phi} * H_3 \big) = -F_1 \wedge F_7 - F_5 \wedge F_3 + C_2 \wedge dF_5 - C_0\, dF_7 \;- \\
- C_2 \wedge H_3 \wedge F_3 + C_0\, H_3 \wedge F_5 + \text{(DBI + WZ)}\;.
\end{multline}
Now we can consistently take $dF_5 - H_3 \wedge F_3 \neq 0$ and $dF_7 - H_3 \wedge F_5 \neq 0$. With a different partial integration we can obtain another equation involving only $C_4$ and $C_6$ and not $C_0$ and $C_2$. Now we can substitute the BI's to get the correct source terms from the bulk action. The result is:
\be
d\big( e^{-\phi} * H_3 \big) = -F_1 \wedge F_7 - F_5 \wedge F_3 - \Big( C_6 - C_4 \wedge \calF + \frac{1}{2} C_2 \wedge \calF^2 - \frac{1}{3!} C_0\, \calF^3 \Big)\wedge \Omega_2 + \dots
\ee
The terms we are still missing are the ones from the DBI and WZ action.

The contribution from the D7-brane action is obtained by recalling that $\delta S_{D7}/\delta B_2 = \delta S_{D7}/\delta \calF$:
\begin{multline}
2\kappa_{10}^2 \frac{\delta S_{D7}}{\delta B_2} = - e^\phi\, \frac{\delta}{\delta\calF}\, \sqrt{-\det(\hat g+ e^{-\phi/2}\, \calF)}\; \delta^{(2)}(D7) \;- \\
- \Big( C_6 - C_4 \wedge \calF + \frac{1}{2} C_2 \wedge \calF^2 - \frac{1}{3!} C_0\, \calF^3 \Big) \wedge \Omega_2 \;.
\end{multline}
Notice that the first piece is not explicitly written as a form. Eventually, summing the bulk and brane contribution to the equation, we get:
\be \label{eom H3}
d\big( e^{-\phi} * H_3 \big) = -F_1 \wedge F_7 - F_5 \wedge F_3 + e^\phi\, \frac{\delta}{\delta\calF}\, \sqrt{-\det(\hat g+ e^{-\phi/2}\, \calF)}\; \delta^{(2)}(D7) \;.
\ee

As we show below, it considerably simplifies in our setup: a spacetime-filling D7-brane in a warped product space, along an holomorphic 4-cycle and with $(1,1)$ anti-self-dual flux. In this case the variation can be written as:
\be \label{DBI variation}
e^\phi\, \frac{\delta}{\delta\calF}\, \sqrt{-\det(\hat g+ e^{-\phi/2}\, \calF)}\; \delta^{(2)}(D7) = -h^{-1}\, \dvol_{3,1} \wedge \calF \wedge \Omega_2 \;.
\ee

% Problematic is also the EOM for $d(e^{-\phi}\ast H_3)$. The bulk computation involves many gauge potentials which are not defined over the D7-brane, so that the source terms are not correctly derived. Consistency requires the bulk contribution to be
% \begin{equation}
% 2\kappa_{10}^2 \frac{\delta S_{bulk}}{\delta B_2} = d(e^{-\phi}\ast H_3) - F_1 \wedge e^\phi\ast F_3 + F_5\wedge F_3 + \Bigl( C_6 - C_4 \wedge \calF + \frac{1}{2} C_2 \wedge \calF^2 - \frac{1}{6} C_0 \calF^3 \Bigr)\wedge \Omega_2 \;.
% \end{equation}
% The contribution from the brane is obtained by recalling that $\delta S_{D7}/\delta B_2 = \delta S_{D7}/\delta \calF$:
% \begin{equation}
% 2\kappa_{10}^2 \frac{\delta S_{D7}}{\delta B_2} = - e^\phi \frac{\delta}{\delta\calF} \sqrt{-\det(\hat g+ e^{-\phi/2}\, \calF)} \delta^{(2)}(D7) - \Bigl( C_6 - C_4 \wedge \calF + \frac{1}{2} C_2 \wedge \calF^2 - \frac{1}{6} C_0 \calF^3 \Bigr)\wedge \Omega_2 \;.
% \end{equation}
% Notice that the first piece is not explicitly written as a form. As I show below, it considerably simplifies in our setup: a spacetime filling D7-brane in a warped product space, along an holomorphic 4-cycle and with $(1,1)$ anti-self-dual flux. Summing the bulk and brane contributions we get:
% \begin{equation} \label{eom H3}
% d\big( e^{-\phi}\ast H_3 \big) = e^\phi \, F_1\wedge\ast F_3 - F_5\wedge F_3 + h^{-1} \text{vol}_{3,1} \wedge \calF\wedge\Omega_2 \;.
% \end{equation}
% Notice that in our setup we have source terms for $F_1$, $F_3$, $F_5$ and $H_7$. Thus $C_6$, $C_8$ and $B_2$ are still well defined gauge potentials, while the other ones are not.

Adding the Wess-Zumino actions for D3 and D5-branes we can in the same way compute the charges they source. In particular:
\begin{equation}
S_{IIB} \supset \mu_3 \int_{D3} \hat C_4 + \mu_5 \int_{D5} \hat C_6
\end{equation}
leads to the effective charges written in equation \eqref{def eff charges} (apart from the factor of 2 mentioned above).

\subsection{$SU(3)$-structure Manifolds and Submanifolds}

I give here some useful formulae. In our setup the D7-brane wraps an holomorphic 4-cycle in a 6d complex $SU(3)$-structure manifold. The 4-cycle inherits a complex structure and a (non-closed) K\"ahler form $\hat J$. Moreover the gauge flux $\calF$ on it is real $(1,1)$ and primitive ($\calF\wedge \hat J=0$). This is equivalent to $\calF = -\ast_4 \calF$ \cite{Gomis:2005wc}. I give an expression for $\sqrt{\det|\hat g + \calF|}$ and its derivatives in this particular case.
\begin{equation} \label{generalized det}
\calF = -\ast_4\calF \qquad\Rightarrow\qquad \sqrt{\det|\hat g + \calF|} \; d^{2n}x = \frac{1}{2} ( \hat J\wedge\hat J - \calF\wedge \calF) \;.
\end{equation}
Moreover in \cite{Gomis:2005wc} it is claimed that
\begin{equation}
\sqrt{\det|\hat g + \calF|} \; d^{2n}x \geq \frac{1}{2} ( \hat J\wedge \hat J - \calF\wedge \calF) \;,
\end{equation}
and the inequality is saturated only for an holomorphic embedding and $\calF = -\ast_4 \calF$.

Then I compute the variation of the determinant under a general variation of $\calF$:
\begin{equation}
\delta\, \sqrt{\det|\hat g+\calF|} = \frac{1}{2} \sqrt{\det|\hat g+\calF|} \, (\hat g + \calF)^{-1\, t \, [ab]} \, \delta \calF_{ab} \;.
\end{equation}
This expression evaluated for an ASD $\calF$ (but still completely general $\delta\calF$) gives:
\begin{equation} \label{variation det}
\delta\, \sqrt{\det|\hat g+\calF|} \; d^4x \Big|_{\calF=-\ast_4\calF} = -\calF\wedge \delta\calF \;.
\end{equation}

\subsection{Probes: SUSY vs EOM's}

With formula \eqref{variation det} at hand it is easy to verify that the $\kappa$-symmetry constraints for the D7-brane imply that the equation of motion of the gauge connection $A$ is satisfied.
Making use of the actual warped product shape of the metric and taking advantage of the $\kappa$-symmetry constraints, the variation of the DBI plus WZ action is evaluated to be
% I write the Dirac-Born-Infeld piece of action as
% \begin{equation}
% S_{D7}^{DBI} = -\mu_7 \int d^8\xi\, e^\phi\, h^{-1}\, \sqrt{\det|\hat g_4 + e^{-\phi/2}\, \calF|} \;,
% \end{equation}
% where I used the explicit warped product shape of our metric. Then the variation of the full D7 action, taking advantage of the $\kappa$-symmetry constraints, is evaluated to be
\begin{equation} \label{sources}
2\kappa_{10}^2 \frac{\delta S_{D7}}{\delta \calF} = - h^{-1} \, \text{vol}_{3,1} \wedge \calF - \hat C_6 + \hat C_4\wedge \calF - \frac{1}{2} \hat C_2 \wedge \calF^2 + \frac{1}{6} \hat C_0\, \calF^3 \;.
\end{equation}
% This is the formula at the origin of \eqref{eom H3}.
The EOM for $A$ states that this must be closed:
\begin{equation}
0 = 2\kappa_{10}^2 \, d\, \frac{\delta S_{D7}}{\delta \calF} = -dh^{-1} \, \text{vol}_{3,1} \wedge \calF - h^{-1}\, \text{vol}_{3,1} \wedge d\calF - \hat F_7 + \hat F_5\wedge \calF - \frac{1}{2} \hat F_3\wedge \calF^2 + \frac{1}{6} \hat F_1\wedge \calF^3
\end{equation}
In our class of solutions the terms $\hat F_3\wedge \calF^2$ and $\hat F_1\wedge \calF^3$ automatically vanish, while the first four terms cancel provided that
\begin{equation}
F_5 = dh^{-1}\wedge \text{vol}_{3,1} + \text{Hodge dual} \qquad\qquad e^\phi\ast_6 F_3 = H_3 \qquad\qquad d\calF = \hat H_3 \;.
\end{equation}
In particular $F_7 = -e^\phi \ast_{10} F_3 = - e^\phi h^{-1} \text{vol}_{3,1}\wedge \ast_6 F_3 = - h^{-1} \text{vol}_{3,1}\wedge H_3$.

\section{SUSY Variations} \label{sec:SUSY variations}

I will make use of the following SUSY variation of Type IIB Supergravity:
\begin{align}
\delta_\epsilon \lambda &= \frac{1}{2} \Gamma^M \Bigl( \partial_M \phi - i\, e^\phi F^{(1)}_M \Bigr) \epsilon + \frac{i}{24} \, e^{\phi/2} \, \Gamma^{MNP} \Bigl( F_{MNP}^{(3)} - i e^{-\phi} H_{MNP}   \Bigr) \epsilon^* \\
\delta_\epsilon \gamma_M &= \partial_M\epsilon + \frac{1}{4} \omega^{NP}_{\phantom{Np}M} \Gamma_{NP} \epsilon + \frac{i}{4} \,e^\phi \, F^{(1)}_M \epsilon + \frac{i}{16\cdot 5!} F^{(5)}_{NPQRS} \Gamma^{NPQRS} \Gamma_M \, \epsilon \\
&\quad - \frac{i}{96} \, e^{\phi/2} \, \Bigl( F_{NPQ}^{(3)} -i e^{-\phi} H_{NPQ}  \Bigr) \Bigl( \Gamma_{M}^{\hspace{1ex}NPQ}-9\delta_M^N\Gamma^{PQ} \Bigr) \epsilon^* \;,
\end{align}
with $\Gamma$'s real matrices. From the metric ansatz \eqref{metric ansatz} we derive the following vielbein, which is compatible with complex structure and orientation:
\begin{equation} \begin{aligned}
e^\mu &= h^{-1/4}\, dx^\mu \\
e^\rho &= h^{1/4} e^u \, d\rho \\
e^\psi &= h^{1/4} \frac{e^u}{3} \, g^5
\end{aligned} \qquad\qquad\qquad
\begin{aligned}
e^{\theta_j} &= h^{1/4} \frac{e^g}{\sqrt{6}} \, d\theta_j \\
e^{\varphi_j} &= - h^{1/4} \frac{e^g}{\sqrt{6}} \, \sin\theta_j\,d\varphi_j \;.
\end{aligned} \end{equation}
The spin connection in vielbein indices is:
\begin{equation} \begin{aligned}
\omega^{\mu\rho} &= - \frac{e^{-u}h'}{4h^{5/4}}\, e^\mu \\
\omega^{\theta_j\rho} &= \frac{e^{-u}(4hg'+h')}{4h^{5/4}}\, e^{\theta_j} \\
\omega^{\varphi_j\rho} &= \frac{e^{-u}(4hg'+h')}{4h^{5/4}}\, e^{\varphi_j} \\
\omega^{\theta_j\varphi_j} &= - \sqrt{6} \, \frac{e^{-g}\cot\theta_j}{h^{1/4}} \, e^{\varphi_j} - \frac{e^{-2g+u}}{h^{1/4}}\, e^\psi
\end{aligned} \qquad\qquad
\begin{aligned}
\omega^{\psi\rho} &= \frac{e^{-u}(4hu'+h')}{4h^{5/4}}\, e^{\psi} \\
\omega^{\psi\theta_j} &= \frac{e^{-2g+u}}{h^{1/4}}\, e^{\varphi_j} \\
\omega^{\psi\varphi_j} &= -\frac{e^{-2g+u}}{h^{1/4}}\, e^{\theta_j} \;.
\end{aligned} \end{equation}

I will perform computations in vielbein indices. From the ansatz for the 5-form flux \eqref{F5 ansatz} we get, contracting with the $\Gamma$'s and lowering the indices:
\begin{equation}
F^{(5)}_{NPQRS} \Gamma^{NPQRS} = 5!\, \frac{e^{-u}h'}{h^{5/4}} ( \Gamma_{0123\rho} + \Gamma_{\psi\theta_1\varphi_1\theta_2\varphi_2}) \;.
\end{equation}

The usual ansatz for the preserved Killing spinor in the bulk is:
\begin{equation}
\begin{aligned}
&\epsilon = h^{-1/8} e^{-i \frac{\psi}{2}} \, \epsilon_0 \\
&\Gamma_{0123\rho\psi\theta_1\varphi_1\theta_2\varphi_2}\, \epsilon = \epsilon
\end{aligned} \qquad\qquad
\begin{aligned}
& \Gamma_{0123}\, \epsilon = -i\, \epsilon \\
& \Gamma_{\rho\psi}\, \epsilon = \Gamma_{\theta_1\varphi_1}\, \epsilon = \Gamma_{\theta_2\varphi_2}\, \epsilon = -i \, \epsilon \;.
\end{aligned}
\end{equation}

We can start computing the variations. The terms containing $\epsilon$ and $\epsilon^*$ give independent equations. Let me start with the $\epsilon$ ones. The dilatino variation gives the equation for $\phi$ in \eqref{geometric system}, while the gravitino variations give the equations for $g$ and $u$. Then we consider the $\epsilon^*$ terms, recall that the $\Gamma$'s are real. The equation we obtain is equivalent to imposing $e^\phi \ast_6 F_3 = H_3$.

I have checked that the solutions satisfy the Bianchi identities and the equations of motion \eqref{modified BI EOM} and \eqref{eom H3} for the form-fields.

\section{Comparison with Ouyang's Procedure} \label{sec:comparison Ouyang}

In \cite{Ouyang:2003df} the same issue as here is addressed: the addition of D7-branes to the Klebanov-Tseytlin background. The author computes the effect of the branes at leading order in $N_f/M$, as a perturbation of the original background. His procedure to extract the first correction to the 3-form flux $G_3$ is imposing the correct $SL(2,\bbZ)$ monodromy as circling around the brane. I would like to briefly comment on how this is equivalent to taking into account the worldvolume gauge bundle, which we saw is so important to obtain the correct scaling $M_{eff} \to M_{eff} - N_f$.

Let me start showing how can a monodromy contain information about localized charges. Let $\tau = C_0 + i e^{-\phi}$ be the axio-dilaton. A D7-brane generates a monodromy $\tau \to \tau + 1$ (where the general $SL(2,\bbZ)$ transformation is $\tau \to \frac{a\tau + b}{c\tau + d}$). Thus
\begin{equation}
1 = C_0 \Bigr|^{2\pi}_0 = \int_\gamma dC_0 = \int_\gamma F_1 = \int_\Sigma dF_1 \qquad\Rightarrow\qquad dF_1 = - \delta_2(D7) \;,
\end{equation}
where $\gamma$ is a 1-cycle circling the D7, $\Sigma$ is a 2-surface having $\gamma$ as a boundary, $\delta_2(D7)$ is a delta-form at the D7 location and $(-)$ comes from the orientation.

What is the information encoded into $G_3 \to \frac{G_3}{c\tau+d}$? In our case $G_3$ has trivial monodromy, so the same is true for $F_3$ and $H_3$. Then:
\begin{equation}
0 = F_3 \Bigr|^{2\pi}_0 = \int_\gamma dF_3 = \int_\gamma (H_3\wedge F_1 + \alpha_4) = \int_\Sigma [ H_3\wedge \delta_2(D7) + d\alpha_4] = - \hat H_3 + \int_\Sigma d\alpha_4 \;.
\end{equation}
The triviality of the monodromy implies that the source term $\alpha_4$ in the Bianchi identity is non-vanishing, and in fact equal to the gauge bundle induced charge:
\begin{equation}
\text{sources} = \alpha_4 = - \calF \wedge \delta_2(D7)
\end{equation}
with $d\calF = \hat H_3$.

\section{Poincar\'e Duals and Exceptional Divisors} \label{sec:exceptional}

On compact oriented manifolds Poicar\'e duality is a canonical isomorphism between $H_p(\calM,\bbR)$ and $H^{n-p}(\calM,\bbR)$, established through the two canonical isomorphisms with $H^{p*}(\calM,\bbR)$ defined using the two linear pairings:
\begin{equation}
(\calC_p,\alpha_p) = \int_{\calC_p} \alpha_p \qquad\text{and}\qquad (\alpha_p, \beta_{n-p}) = \int \alpha_p \wedge \beta_{n-p} \;.
\end{equation}
Equivalently, the duality $\calC_p \leftrightarrow \omega_{n-p}$ can be established requiring that for every cohomology class $\alpha_p$:
\begin{equation}
\int_{\calC_p} \alpha_p = \int \alpha_p \wedge \omega_{n-p} \;.
\end{equation}
Given a metric, one can also define Hodge duality from $H^p(\calM,\bbR)$ to $H^{n-p}(\calM,\bbR)$. Poicar\'e duality maps the intersection operator $\cap$ in homology to the wedge operator $\wedge$ in cohomology. If the dimension of $\calM$ is $n=2l$ then the intersection number is given by
\begin{equation}
\# (C_l,D_l) = \int_{C_l \cap D_l} 1 = \int \omega_l\wedge \sigma_l \;.
\end{equation}

In order to understand the geometry and the induced charges of probe branes at the conifold singularity it is better to resolve it. This process in general breaks supersymmetry, but it is a good way of computing topological quantities such as charges. The metric and the K\"ahler form of the resolved conifold are \cite{Candelas:1989js}:
\begin{align}
ds_6^2 &= \frac{1}{k(r)} dr^2 + \frac{r^2}{9} k(r) (g^5)^2 + \frac{r^2}{6} (d\theta_1^2 + \sin^2\theta_1\,d\varphi_1^2) + \frac{r^2+a^2}{6} (d\theta_2^2 + \sin^2\theta_2\,d\varphi_2^2) \\
J &= \frac{r}{3} dr\wedge g^5 - \frac{r^2}{6} \sin\theta_1\,d\theta_1 \wedge d\varphi_1 - \frac{(r^2+a^2)}{6} \sin\theta_2\,d\theta_2 \wedge d\varphi_2 \\
k &= \frac{r^2+9a^2}{r^2+6a^2} \;.
\end{align}
The coordinates have range: $r\in [0,\infty)$, $\psi \in [0, 4\pi)$, $\theta_j \in [0,\pi]$ and $\varphi_j \in [0,2\pi)$. In this appendix I will care attention to minus signs in $J$ which reappear in the pulled-back volume forms in 4d and 2d, and consequently in the computation of integrals.

Consider the two non-compact 4-cycles $\Sigma_j = \{\theta_j,\varphi_j = \text{const} \}$. From the expression of the metric it is easy to see that $\Sigma_1$ has a non-vanishing 2-cycle at the origin and thus it is $\widehat{\bbC^2}$ blown up at a point. Of course the 2-cycle is exactly the same as the one blown up to resolve the conifold. Instead $\Sigma_2$ still has the topology of $\bbC^2$ and only touches the 2-cycle at a point. Under a flop transition the r\^ole of the two 4-cycles gets exchanged.

I construct a resolved $B_2$ on the resolved conifold following the requirements: $B_2$ is $(1,1)$, closed and primitive ($B_2\wedge J\wedge J=0$). I start with an ansatz constructed taking the three pieces of $J$ with general functions $f_{i=1,2,3}(r)$ in front. Primitivity fixes the relation $f_1 = f_2 + f_3$. Closure gives us a system of two linear first order ODE's. Only one of the two solutions is regular at the origin:
\begin{equation}
B_2 = \frac{\pi\, b_0}{2} \Bigl\{ - \frac{2ra^2}{(r^2+a^2)^2} dr\wedge g^5 + \frac{r^2}{r^2+a^2}\, \sin\theta_1\,d\theta_1 \wedge d\varphi_1 - \frac{r^2+2a^2}{r^2+a^2}\, \sin\theta_2\,d\theta_2 \wedge d\varphi_2 \Bigr\} \;.
\end{equation}
The normalization is fixed such that $\int_{S^2} B_2 = 4\pi^2 b_0$, where $S^2 = \{\theta_1=\theta_2,$ $\varphi_1=-\varphi_2;$ $r,\psi=\text{const}\}$. Notice that $B_2$ approaches a constant non-zero value at infinity. This is because the geometry has a 2-cycle supporting it.

Now I go on with the construction of $\calF = \hat B_2 + 2\pi F_2$ on the 4-cycles of interest. I am looking for fluxes that fall off at infinity, because in the singular limit I only want finite induced charges. Consider $\Sigma_2$, with topology of $\bbC^2$. Not there being any 2-cycle we can simply set $F_2$ to cancel $\hat B_2$, so that $\calF|_{\Sigma_2} = 0$. On $\Sigma_1$ with topology of $\widehat{\bbC^2}$ the situation is different. We cannot set $2\pi F_2$ equal and opposite to $\hat B_2$, because its flux is quantized on $S^2$. We can instead set a closed $F_2$ with vanishing flux on $S^2$ that kills the tail of $\hat B_2$:
\begin{equation} \label{flux resolved sigma1}
\calF = \hat B_2 + 2\pi F_2 = \frac{\pi\, b_0}{2} \Bigl\{ - \frac{4ra^2}{(r^2+a^2)^2} dr\wedge \hat g^5 - \frac{2a^2}{r^2+a^2}\, \sin\theta_2\,d\theta_2 \wedge d\varphi_2 \Bigr\} \;.
\end{equation}
One can explicitly verify that $\calF \wedge \hat J = 0$, $\int_{S^2} \calF = 4\pi^2 b_0$ and $\int \calF\wedge\calF = - (4\pi^2 b_0)^2$. On the other hand, with a different choice of $F_2$ we could also add further flux on $S^2$, obtaining the same $\calF$ of \eqref{flux resolved sigma1} but with $b_0 \to b_0 + \phi_0$.

Such an $\calF$ is in fact proportional to the anti-self-dual (and primitive) Poincar\'e dual of $S^2$ on $\Sigma_1$. The two integrals tell us that the self-intersection number of $S^2$ in $\widehat{\bbC^2}$ is $-1$. This is true in general: the exceptional $S^2$ arising in the blowing up of a smooth point has self-intersection number $-1$.

I would like to conclude with the 4-cycle $\Sigma_K = \{ \theta_1 = \theta_2, \varphi_1 = \varphi_2 \}$ which has the topology of $\widehat{\bbC^2/\bbZ_2}$ blown up at the origin. In this case $\hat B_2$ falls off at infinity (indeed in the singular limit $\hat B_2=0$) but $\hat B_2 \wedge \hat J \neq 0$ so that again we need to add a suitable fluxless $F_2$.
Again $\hat B_2$ is proportional to the Poincar\'e dual to the 2-cycle, even if it is not the anti-self-dual representative in the cohomology class. One can compute $\int_{S^2} B^2 = 4\pi^2 b_0$ and $\int_{\Sigma_K} B_2\wedge B_2 = - \frac{1}{2} (4\pi^2b_0)^2$, confirming that the self-intersection number of $S^2$ in $\Sigma_K$ is $-2$. A generic $\bbC^2/\bbZ_N$ singularity is resolved by blowing up $N-1$ intersecting $\bbP^1$'s $E_j$, whose non-vanishing intersections are
\begin{equation}
E_i \cdot E_i = -2 \qquad\qquad E_i \cdot E_{i+1} = 1 \;.
\end{equation}

\

The curvature couplings appearing in the reduced D7 WZ actions \eqref{reduced D7 action C2} and \eqref{reduced D7 action C2 hat} are obtained by explicit evaluation of
\begin{equation}
(\text{curv}) = - \frac{1}{48} \int_\Sigma p_1(\calR) = \frac{1}{48} \int_\Sigma \frac{\Tr \, \calR\wedge\calR}{8\pi^2} \;,
\end{equation}
and on a four dimensional manifold we have:
\begin{equation}
\Tr \, \calR\wedge\calR = \frac{1}{4} R^a_{\phantom{a}b\,cd} \, R^b_{\phantom{b}a\,ef} \, \epsilon^{cdef} \, d\xi^{1234} \;.
\end{equation}
In fact in non-compact cases we cannot appeal to its relation to topological invariants, as the behavior of the metric at infinity influences the result.%
\footnote{I thank G.~Bonelli for explanations on that.}
I found:
\begin{equation}
\bbC^2: \quad (\text{curv}) = \frac{1}{216} \qquad\qquad \widehat{\bbC^2}: \quad (\text{curv}) = - \frac{23}{432} \;.
\end{equation}
It would be nice to understand the meaning of these numbers.

\end{document}